\documentclass[10pt]{amsart}
\usepackage{amsmath,amssymb,amscd,graphicx}
\usepackage{lpic}
\usepackage{hyperref}

\newtheorem{theorem}{Theorem}

\theoremstyle{definition}

\theoremstyle{remark}

\allowdisplaybreaks \numberwithin{equation}{section}

\newcommand{\PP}{{\mathbb P}}
\def\Im{\mathop{\rm Im}\nolimits}
\def\Re{\mathop{\rm Re}\nolimits}

\def\Jac{\mathop{\rm Jac}\nolimits}

\def\de{\mathrm{d}}
\def\BH{\widehat{{B}}}

\def\abelmap{\mathcal{A}}
\def\aq{\mathfrak{a}}
\def\bq{\mathfrak{b}}
 \newcommand{\ah}{\hat{\mathfrak{a}}}
 \newcommand{\bh}{\hat{\mathfrak{b}}}
 \newcommand{\azero}{\hat{\mathfrak{a}}^{{0}}}
 \newcommand{\bzero}{\hat{\mathfrak{b}}^{{0}}}

\newcommand{\cred}[1]{\color{red}{#1}}
\newcommand{\cblu}[1]{\color{blue}{#1}}
\newcommand{\enf}[1]{\textbf{\textit{#1}}}
\begin{document}
\title{On Charge-3 Cyclic Monopoles}
\author{H.~W. Braden}
\address{School of Mathematics, Edinburgh University, Edinburgh.}
\email{hwb@ed.ac.uk}
\author{Antonella D'Avanzo}
\address{School of Mathematics, Edinburgh University, Edinburgh.}
\email{a.davanzo@ed.ac.uk}
\author{V.~Z. Enolski}
\address{Institute of Magnetism, National Academy of Sciences of
Ukraine.} \email{vze@ma.hw.ac.uk}
\date{\today}
 \maketitle

 \begin{abstract}
We determine the spectral curve of charge 3 BPS $su(2)$ monopoles
with $\texttt{C}_3$ cyclic symmetry. The symmetry means that the
genus 4 spectral curve covers a (Toda) spectral curve of genus 2.
A well adapted homology basis is presented enabling the theta
functions and monopole data of the genus 4 curve to be given in
terms of genus 2 data. The Richelot correspondence, a
generalization of the arithmetic mean, is used to solve for this
genus 2 curve. Results of other approaches are compared.
 \end{abstract}

\tableofcontents

\section{Introduction}
The first order Bogomolny equations,
\begin{equation}
\label{bpsbogomolny}
B_i=\frac{1}{2}\sum_{j,k=1}\sp3\epsilon_{ijk}F\sp{jk}=D_i\Phi,
\end{equation}
are rather ubiquitous. They arose while studying a limit of
Yang-Mills-Higgs gauge theory in three space dimensions in which
the the Higgs potential is removed but a remnant of this remains
in the boundary conditions associated with (\ref{bpsbogomolny}).
Here $F_{ij}$ is the field strength associated to a gauge field
$A$, and $\Phi$ is the Higgs field. The same equations may also be
viewed as a dimensional reduction of the four dimensional
self-dual equations upon setting all functions independent of
$x_4$ and identifying $\Phi=A_4$; they are also encountered in
supersymmetric theories when requiring certain field
configurations to preserve some fraction of supersymmetry. Just as
the self-duality equations admit instanton solutions in four
dimensions, the Bogomolny equations possess topological soliton
solutions with particle-like properties, known as magnetic
monopoles, and these have been the subject of considerable
interest over the years \cite{ms04}. Early on a curve
$\hat{\mathcal{C}}$ was found to be associated to these BPS
monopoles. Indeed the same curve arose by (at least) two different
routes both with origins in instanton theory. Whilst considering
the Atiyah-Ward instanton ansatz in the monopole setting Corrigan
and Goddard \cite{cg81} encountered $\hat{\mathcal{C}}$ and this
was given a twistorial description by Hitchin \cite{hitchin82}.
Just as Ward's twistor transform relates instanton solutions on
$\mathbb{R}\sp4$ to certain holomorphic vector bundles over the
twistor space $\mathbb{CP}\sp3$, Hitchin showed that the
dimensional reduction leading to BPS monopoles could be made at
the twistor level as well and the curve $\hat{\mathcal{C}}$
naturally lies in mini-twistor space, $\hat{\mathcal{C}}\subset$
T$\mathbb{P}\sp1$. The second appearance of the curve
$\hat{\mathcal{C}}$ is closely connected with integrable systems.
Nahm gave a transform of the ADHM instanton construction to
produce BPS monopoles \cite{nahm82} and the resulting Nahm's
equations have Lax form with corresponding spectral curve
$\hat{\mathcal{C}}$. Hitchin \cite{hitchin83} proved that all
monopoles could be obtained by Nahm's approach provided the curve
$\hat{\mathcal{C}}$ was subject to certain nonsingularity
conditions. Bringing methods from integrable systems to bear upon
the construction of solutions to Nahm's equations for the gauge
group $SU(2)$ Ercolani and Sinha \cite{es89} later showed how one
could solve (a gauge transform of) the Nahm equations in terms of
a Baker-Akhiezer function for the curve $\hat{\mathcal{C}}$. Thus
given a curve $\hat{\mathcal{C}}$ the machinery of integrable
systems allows one (in principle)  to construct solutions to
Nahm's equations and thence monopoles \cite{bren09a}.

The problem in the approach just described, and to which this
paper is devoted, is in constructing the curve
$\hat{\mathcal{C}}$: some of the conditions necessary for the
regularity of solutions just alluded to impose transcendental
constraints on $\hat{\mathcal{C}}$, and we presently lack analytic
means for solving these. One such constraint comes about by
requiring the periods of a meromorphic differential on
$\hat{\mathcal{C}}$ to be specified. This type of constraint
arises in many other settings as well, for example when specifying
the filling fractions of a curve in the AdS/CFT correspondence
\cite{kmmz}, finding closed geodesics on an ellipsoid \cite{af06}
or constructing harmonic maps  $ T ^ 2\rightarrow S ^ 3 $
\cite{hitchin90}. The second type of constraint is that the linear
flow on the Jacobian of $\hat{\mathcal{C}}$ corresponding to the
integrable motion only intersects the theta divisor in a
prescribed manner; equivalently this may be expressed as the
vanishing of a real one parameter family of cohomologies of
certain line bundles on $\hat{\mathcal{C}}$. While techniques
exist that count the number of intersections of a complex line
with the theta divisor we are unaware of anything comparable in
the real setting \cite{bren09}. Thus the application of integrable
systems techniques to the construction of monopoles (and indeed
more generally) encounters two types of problem that each merit
further study.

In the present paper we will simplify then solve these constraints
by imposing spatial symmetries on the monopole. Imposing symmetry
reduces the number of constraints to be solved for and here we
shall focus on charge $3$ monopoles with cyclic symmetry
$\texttt{C}_3$.  The spectral curve of a charge $n$ monopole may
be expressed in the form
\begin{equation*}
P(\eta,\zeta):=\eta^n+\eta^{n-1} a_1(\zeta)+\ldots+\eta^r
a_{n-r}(\zeta)+ \ldots+\eta\,
a_{n-1}(\zeta)+a_n(\zeta)=0,\label{spectcurve}
\end{equation*}
where $a_r(\zeta)$  (for $1\leq r\leq n$) is a polynomial in
$\zeta$ of maximum degree $2r$. Hitchin's construction involves
three constraints on the curve. The first (\textbf{H1}) requires
the curve $\hat{\mathcal{C}}$ to be real with respect to the
standard real structure on $T\PP\sp1$,
\begin{equation}\label{realinv}
\tau:(\zeta,\eta)\mapsto(-\frac{1}{\bar{\zeta}},
-\frac{\bar{\eta}}{\bar{\zeta}^2}),
\end{equation}
the anti-holomorphic involution defined by reversing the
orientation of the lines in ${\mathbb R}\sp3$. As a  consequence
the coefficients of the curve satisfy
$a_r(\zeta)=(-1)^r\zeta^{2r}\overline{a_r(-{1}/{\overline{\zeta}})}
$. Imposing spatial symmetries via fractional linear
transformations of $T\PP\sp1$ simplifies the curve. Long ago
monopoles of charge $n$ with cyclic symmetry $\texttt{C}_n$ were
shown to exist \cite{or82} and these correspond to curves
invariant under $(\eta,\zeta)\rightarrow (\omega\eta,\omega\zeta)$
(where $\omega=\exp(2i\pi /n)$). Imposing Hitchin's reality
conditions and centering the monopole (setting $a_1=0$) then gives
us the $\texttt{C}_n$ symmetric spectral curve in the form
\begin{equation*}\label{cycliccurve}\eta\sp{n}+a_2
\eta\sp{n-2}\zeta\sp2+\ldots+a_n\zeta\sp{n}+\beta\zeta\sp{2n}+(-1)\sp{n}\bar\beta=0,\qquad
a_i\in\mathbf{R}. \end{equation*} By an overall rotation we may
choose $\beta$ real and so the charge 3 spectral curves
$\hat{\mathcal{C}}$ we will focus on in this paper have the form
\begin{equation}
\eta^3+\alpha\eta\zeta^2+\beta\zeta^6+\gamma
\zeta^3-\beta=0.
\label{curve}
\end{equation}
The remaining two constraints of Hitchin on this curve are the
transcendental constraints referred to above. The first of these
(\textbf{H2}) may be expressed in the following manner
\cite{hmr99}: given a canonical homology basis
$\{\hat{\mathfrak{a}}_\mu,\hat{\mathfrak{b}}_\mu\}$
for the curve $\hat{ \mathcal{C}}$ there exists a 1-cycle
$\widehat{\mathfrak{es}}=\boldsymbol{n}\cdot{\hat{\mathfrak{a}}}+
\boldsymbol{m}\cdot{\hat{\mathfrak{b}}}$ such that for every
holomorphic differential
\begin{equation}\Omega=\dfrac{\beta_0\eta^{n-2}+\beta_1(\zeta)\eta^{n
-3}+\ldots+\beta_{n-2}(\zeta)}{{\partial\mathcal{P}}/{\partial
\eta}}\,d\zeta, \qquad
\oint\limits_{\widehat{\mathfrak{es}}}\Omega=-2\beta_0.
\label{eshit2} \end{equation} Dually the vector
$$\boldsymbol{\widehat U}= \frac12
\boldsymbol{n}+\frac12\hat{\tau}\boldsymbol{m} $$ (where
$\hat{\tau}$ is the period matrix of $\hat{ \mathcal{C}}$)
is a half-period. The vector $\boldsymbol{\widehat U}$ is known as
the Ercolani-Sinha vector \cite{es89} and may be expressed as the
periods of a meromorphic differential. These conditions, known as
the Ercolani-Sinha constraints, impose $(n-1)^2$ (the genus of
$\hat{\mathcal{C}}$) \emph{transcendental constraints} on the
curve. The remaining constraint (\textbf{H3}) is that for a
special vector $\widetilde{\boldsymbol{K}}$ in the Jacobian the
linear flow $\lambda\boldsymbol{\widehat
U}-\widetilde{\boldsymbol{K}}$ intersects the theta divisor only
at $\lambda=0$ and $2$. The consequences of assuming symmetry is
that the spectral curve covers another curve, the quotient curve
by the symmetry. In the case of cyclic symmetry we have an
$n$-fold unbranched cover $\pi:\hat{\mathcal{C}}\rightarrow
\mathcal{C}$ of a hyperelliptic curve of genus $n-1$ which is the
spectral curve of the $a_n$ affine Toda system, and \cite{bra10}
shows how both the constraints (\textbf{H2,3}) reduce to become
constraints on the reduced curve. In particular
$$\lambda\widehat{\boldsymbol{U}}-
\widetilde{\boldsymbol{K}}=
\pi\sp\ast(\lambda \boldsymbol{U} -{\boldsymbol{K}}_{\infty_+}+e).
$$
where $\boldsymbol{U}$ expresses the Ercolani-Sinha constraints of
the curve $\mathcal{C}$ and the other quantities will be defined
later. Thus in this paper we have a 3-fold unbranched cover of the
of the curve $\mathcal{C}$ given by
\begin{equation}\label{quotientcurve}
y^{2}=(x^{3}+\alpha x +\gamma)^{2}+4\beta^{2}.
\end{equation}

To solve the remaining transcendental constraints for the reduced
curve $\mathcal{C}$ we now use several pieces of research. First
the work of \cite{bren06, bren09}\footnote{The first of these
papers consisted of two parts that have been separately published
as \cite{bren10a, bren10b}.} identifies the solutions of
(\ref{eshit2}) for the class of curves (\ref{curve}) with
$\alpha=0$ and \cite{bren09} shows that the only solutions of the
Hitchin constraints are for the curves
\begin{equation}\label{tmcurve}
\eta^3+\chi(\zeta^6\pm 5\sqrt{2}
\zeta^3-1)=0, \qquad \chi^{\frac{1}{3}} =-\frac16\,
    \frac{\Gamma(\frac16)\Gamma(\frac13)}{2\sp\frac16\,
    \pi\sp{\frac12}}.
\end{equation}
These correspond to tetrahedrally symmetric monopoles and for each
sign of $\pm 5\sqrt{2}$  there is a unique vector
$\boldsymbol{\widehat U}$. Next we use the work of \cite{hmm95}.
Here cyclically symmetric monopoles (and more generally, those
with Platonic spatial symmetries) were reconsidered from a variety
of perspectives. Cyclically symmetric monopoles form
$4$-dimensional totally geodesic submanifolds
$\mathcal{M}_n\sp{l}$ of the full moduli space of charge $n$
monopoles, where $0\le l<n$. Ignoring the rotational degrees of
freedom these then yield one dimensional submanifolds. By
considering the rational map description of these monopoles
Hitchin, Manton and Murray were able to further specify the
$\mathcal{M}_n\sp{l}$ which may be viewed as orbits of geodesic
monopole scattering. For charge $3$ there were five loci of
spectral curves of the form (\ref{curve}). Of these loci, four
were isomorphic: at one end asymptotically one has
$\alpha^3=27\beta^2$ (with $\beta$ of either sign) and $\gamma=0$
while at the other end $\alpha=\pi^2/4-3b^2$, $\beta=0$ and
$\gamma=2b(b^2+\pi^2/4)$ (with $b$ of either sign). Half-way along
this is the tetrahedrally symmetric monopole, the four loci
corresponding to four distinct orientations of the tetrahedron.
The final locus corresponds to the family of curves with
$\gamma=0$ and where the symmetry is enlarged to the dihedral
symmetry $\texttt{D}_3$: asymptotically we have
$\alpha^3=27\beta^2$ (with $\beta$ large and positive at one end
and negative at the other) and half-way along this there is the
axisymmetric monopole. We use this work as follows. Because the
Ercolani-Sinha vector $\boldsymbol{\widehat U}$ is discrete, this
will be constant for each of the loci emanating from the
tetrahedrally symmetric points. Starting then at a point
$(\alpha,\beta,\gamma)$ corresponding to a tetrahedrally symmetric
monopole we deform away from this by solving the (reduced)
Ercolani-Sinha constraint for the given fixed
$\boldsymbol{\widehat U}$. Thus we will obtain the loci
$\gamma\ne0$. This idea is similar to that used by Sutcliffe
\cite{sut96b} when obtaining numerical approximations to
(\ref{curve}) by analysing the Nahm equations and using the fact
that the tetrahedral monopole was on one of the loci. In deforming
from the tetrahedrally symmetric points we will use a genus 2
variant of the arithmetic-geometric mean (AGM) (that will be
described more fully in the sequel). Although this deformation is
defined by the (reduced) constraint (\textbf{H2}) the ensuing loci
must also satisfy (\textbf{H3}) for dimensional reasons. Thus we
arrive at the spectral curves (\ref{curve}) with $\gamma\ne0$ that
describe $\texttt{C}_3$ symmetric monopoles. We remark that the
$\gamma=0$ spectral curves further cover an elliptic curve and
these are amenable to a different analysis that will be given
elsewhere.

An outline of the paper is as follows. In section 2 we study the
curves (\ref{curve}, \ref{quotientcurve}) in some detail
determining those quantities needed to reconstruct the
Baker-Akhiezer functions of the integrable systems approach.
Critical here is determining an homology basis that reflects well
the symmetries of the curve $\hat{\mathcal{C}}$. Such a basis both
relates and simplifies the forms of the period matrices of both
$\hat{\mathcal{C}}$ and ${\mathcal{C}}$; it also reduces the
numbers of periods to be calculated to construct the full period
matrices. Perhaps the nicest feature of this homology basis (and
that induced on ${\mathcal{C}}$) is that it enables us to make use
of a remarkable factorisation theorem due to Accola and Fay
\cite{acc71, fay73} and also observed by Mumford. This allows the
theta functions of $\hat{\mathcal{C}}$ to be described in terms of
the theta functions of ${\mathcal{C}}$ for the parts of the
Jacobian that are relevant for us \cite{bra10}. By the end of
section 2 we have reduced the construction of cyclically invariant
monopoles (with $\gamma\ne0$) to questions about a genus two
hyperelliptic curve $\mathcal{C}$. Section 3 then discusses the
restrictions the Ercolani-Sinha constraints place on
$\mathcal{C}$. The Ercolani-Sinha constraints are shown to reduce
to the single constraint $\int_{\boldsymbol{\mathfrak{c}}}dX/Y=0$
on a scaled form of $\mathcal{C}$, $Y^2=(X^3+a\, X+g)^2+4$ (where
$(a,g):=(\alpha/\beta\sp{2/3},\gamma/\beta)$), and here
$\boldsymbol{\mathfrak{c}}:=\pi(\widehat{\mathfrak{es}})$ is
known. Thus the problem has become one of understanding the
periods of this (scaled) genus two curve as a function of $(a,g)$
with the Ercolani-Sinha yielding $g=g(a)$. We will solve this
transcendental constraint numerically using a genus two variant of
the arithmetic-geometric mean due to Richelot. Section 4 recalls
this theory and describes an extension needed for the curves
relevant here, which have complex conjugate branchpoints. Section
5 then implements this. We conclude with a discussion.

\section{The curve}

In this section we consider the curve $\hat{\mathcal{C}}$
(\ref{curve}) and the quotient curve $\mathcal{C}$ in more detail.
After describing the curves we shall construct homology bases that
enables several simplifications. In particular both the period
matrices and vectors of Riemann constants will be described for
these bases. Throughout we will set $\rho=\exp({2}i\pi/3)$.

\subsection{Branchpoints and monodromy} \label{properties}
\begin{figure}
 \hspace*{-6mm}
\includegraphics[height=250pt]{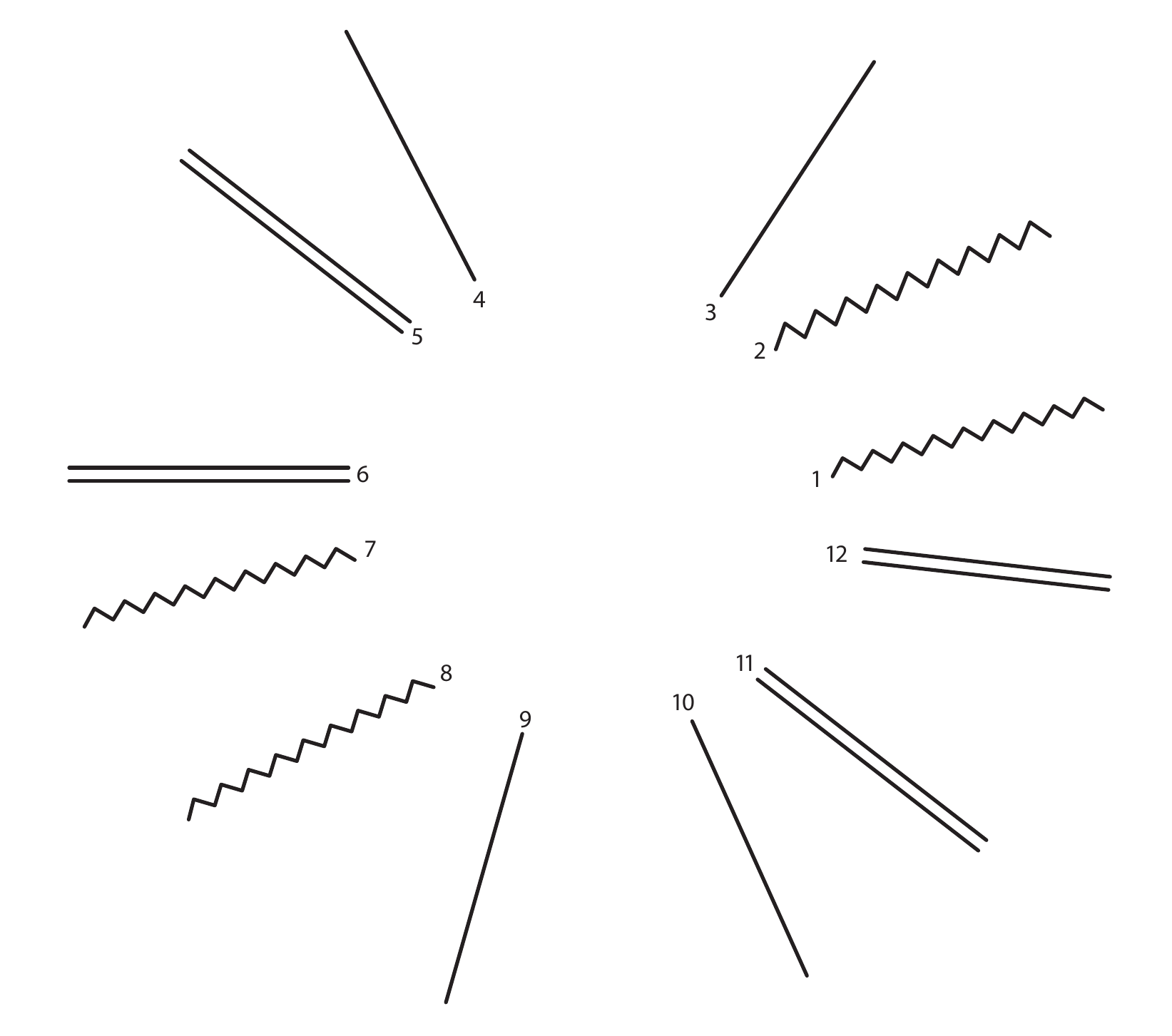}
\hspace*{-5mm}
\begin{minipage}{100pt}\vspace*{-9mm}
\includegraphics[height=80pt]{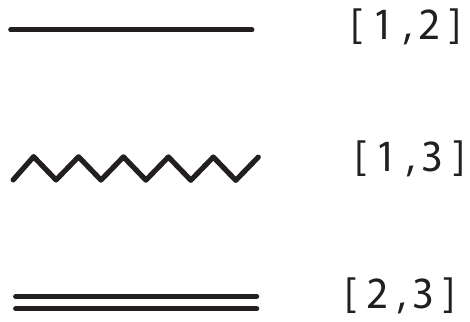}
\end{minipage}
\caption{Branchpoints and monodromy for $\hat{\mathcal{C}}$ }\label{bpts4}
\end{figure}

 The curve (\ref{curve}) has genus 4 and
is not hyperelliptic. A basis for the holomorphic differentials
may be taken to be
\begin{equation}\label{diffxhat}
{\hat{\mathbf{u}}_{1}}=\frac{\de \zeta}{3 \eta^{2}+\alpha \zeta^{2}}, \qquad
{\hat{\mathbf{u}}_{2}}=\frac{\zeta\de \zeta}{3 \eta^{2}+\alpha \zeta^{2}}, \qquad
{\hat{\mathbf{u}}_{3}}=\frac{\zeta^{2}\de \zeta}{3 \eta^{2}+\alpha \zeta^{2}}, \qquad
{\hat{\mathbf{u}}_{4}}=\frac{\eta\de \zeta}{3 \eta^{2}+\alpha \zeta^{2}}.
\end{equation}
Our curve $\hat{\mathcal{C}}$  may be viewed as a 3 sheeted cover
of $\mathbb{P}\sp1$ with 12 ramification points, whose
$\zeta$-coordinates are given by
\begin{align}\label{bptsexp}
\BH^{\zeta}_{1,2}&=\frac{1}{18\beta}\left( -9\gamma-2i\sqrt{3}\alpha^{3/2}\pm
\Delta_{+}^{1/2} \right), \nonumber \\
\BH^{\zeta}_{3,4}&=\frac{1}{18\beta}
\left( -9\gamma+2i\sqrt{3}\alpha^{3/2}\pm \Delta_{-}^{1/2} \right)\\
\BH_{j+4k}^\zeta&=\rho^k (\BH^{\zeta}_{j})^{1/3},\quad j=1,\ldots,4,\; k=0,1,2 , \quad \nonumber
\end{align}
where
$$\Delta_{\pm}= 324\beta^2-3(2\alpha^{3/2}\pm 3i\sqrt{3}\gamma)^2.$$
In Figure \ref{bpts4} we
give a qualitative sketch of the branchpoints; the general
properties of $\hat{\mathcal{C}}$ do not change with the
parameters unless $\alpha=0$, which is a degenerate case that will
be examined separately. The monodromy around each branch point is
found to be
\begin{equation*}
\BH_{{1}},\; \BH_{{2}}, \;\BH_{{7}}, \;\BH_{{8}}\longrightarrow  [1,3],\quad
\BH_{{3}},\; \BH_{{4}}, \;\BH_{{9}},\; \BH_{{10}}\longrightarrow  [1,2],\quad
\BH_{{5}},\; \BH_{{6}}, \;\BH_{{11}},\, \BH_{{12}}\longrightarrow  [2,3].
\end{equation*}

In addition to the real involution (\ref{realinv}) and cyclic
symmetry
\begin{align}\sigma: \; (\zeta,\eta)&\rightarrow(\rho \zeta,\rho \eta)
\label{cyclic}\intertext{the curve possesses the inversion symmetry}
\phi:(\zeta,\eta)&\rightarrow \left(-\dfrac{1}{{\zeta}}, -\dfrac{{\eta}}{{\zeta}^{2}}\right).\label{inv}
\end{align}
The branchpoints (\ref{bptsexp}) form four orbits under the cyclic
symmetry according to
\begin{align*}
\BH_{1}\;\overset{\sigma}{\longrightarrow}&\;\BH_{5}\;\overset{\sigma}{\longrightarrow}\BH_{9},\qquad
&\BH_{2}\;\overset{\sigma}{\longrightarrow}\;\BH_{6}\;\overset{\sigma}{\longrightarrow}\BH_{10},\\
\BH_{3}\;\overset{\sigma}{\longrightarrow}&\;\BH_{7}\;\overset{\sigma}{\longrightarrow}\BH_{11},\qquad
&\BH_{4}\;\overset{\sigma}{\longrightarrow}\;\BH_{8}\;\overset{\sigma}{\longrightarrow}\BH_{12}.
\end{align*}

\begin{figure}
\begin{center}
 \includegraphics[height=220pt]{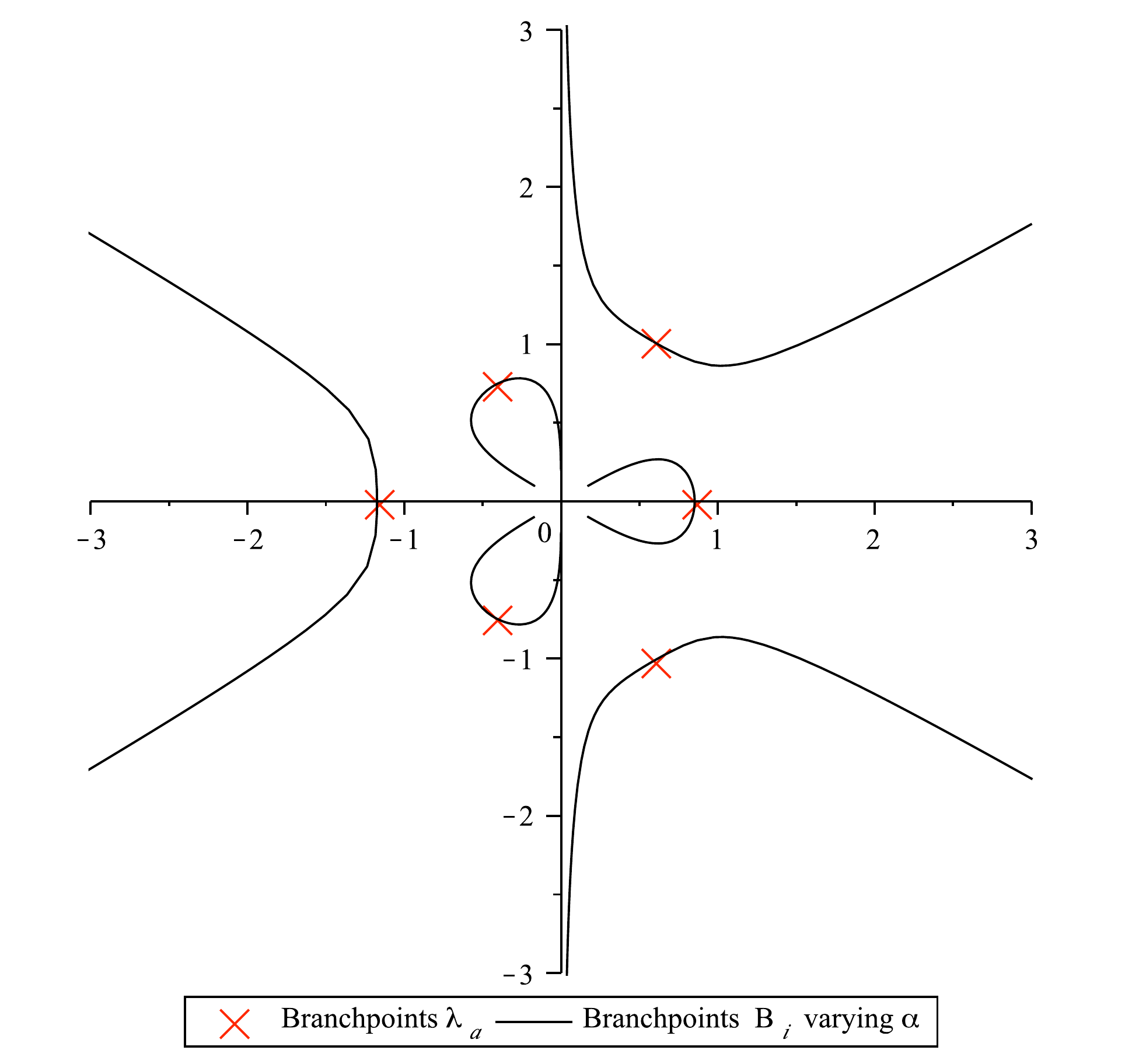}
\caption{Branchpoints for $\alpha\to 0$}\label{alphato0}
\end{center}
\end{figure}

\begin{figure}
\begin{center}
\begin{minipage}{160pt}\vspace*{-2mm}\hspace*{-7mm}
 \includegraphics[width=190pt]{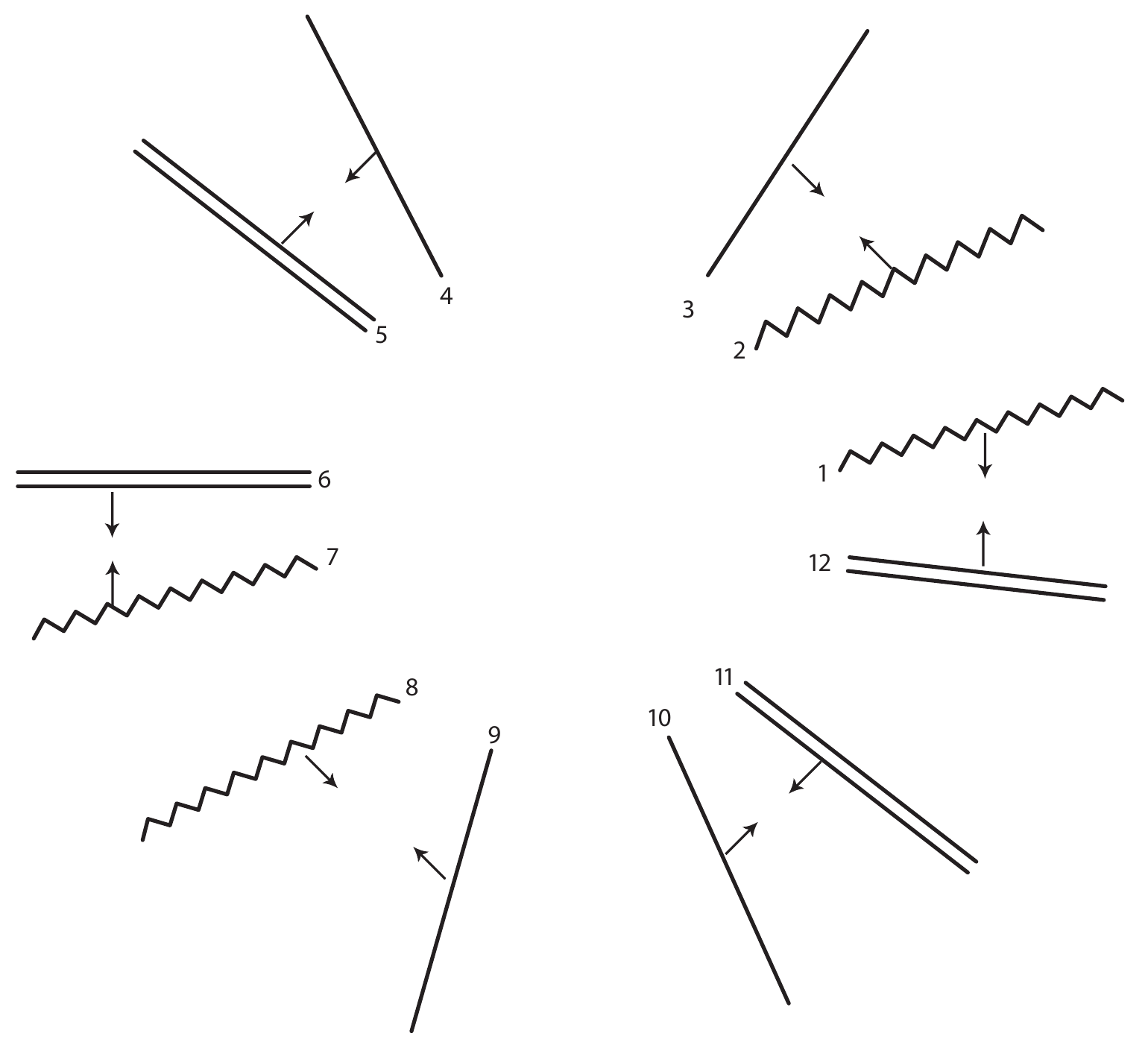}
\end{minipage} 
\begin{minipage}{160pt}\vspace*{-7mm}
\includegraphics[width=190pt]{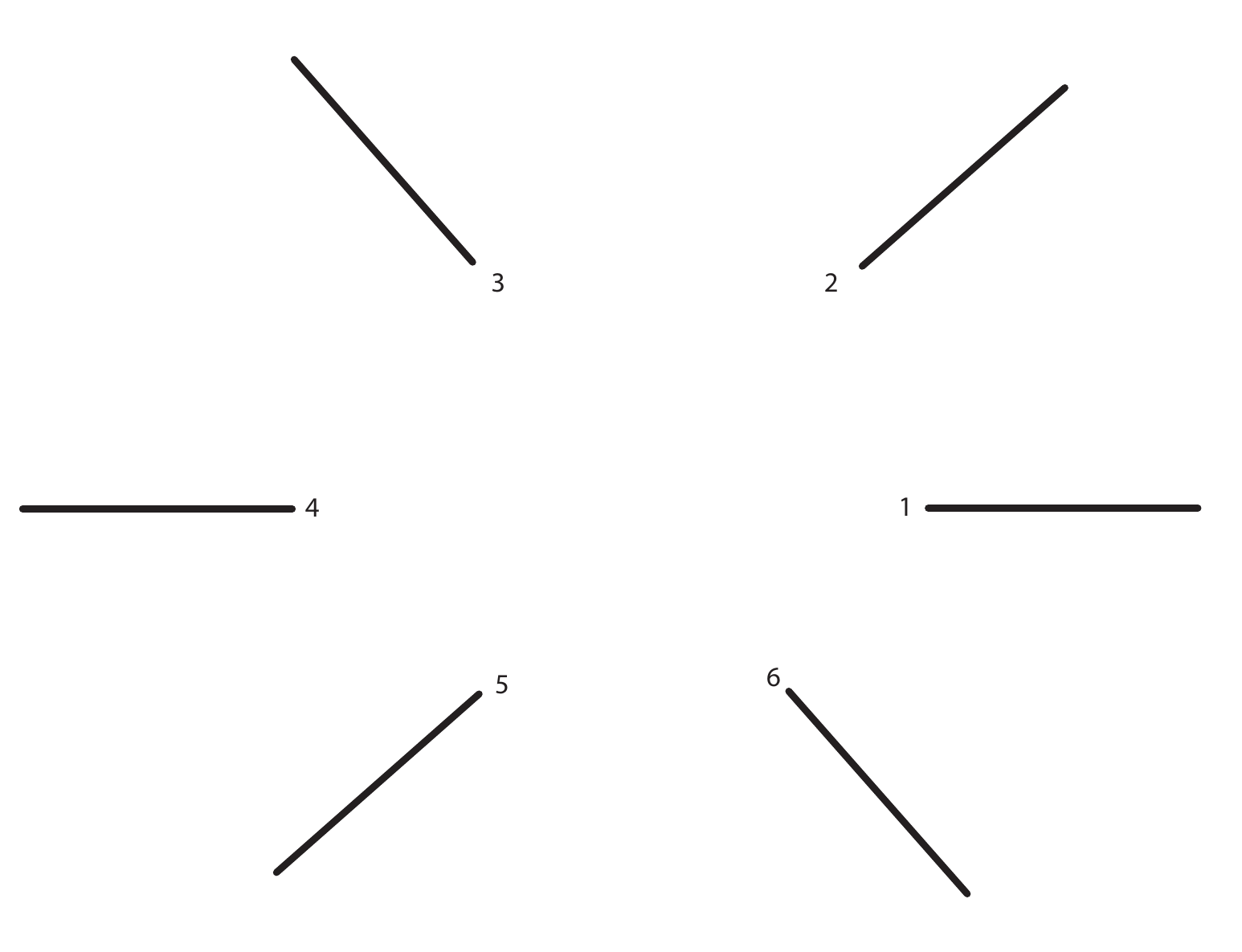}
\end{minipage}
\caption{Branchpoints and monodromy for $\alpha\rightarrow0$.}\label{bpts0}
\end{center}
\end{figure}

\noindent{\emph{The case $\alpha=0$.}} The case $\alpha=0$  is the
curve studied in \cite{bren06}.  The corresponding Riemann surface
also has  genus 4, but now only six branchpoints  $\lambda_{i}$ ($
i=1\ldots 6$),
\begin{align*}\label{lambda}
\lambda_{1}&=\frac{1}{6\beta}\left( -3\gamma + \frac{1}{3}\Delta^{1/2} \right),
&\qquad \lambda_{4}&=\frac{1}{6\beta}\left( -3\gamma - \frac{1}{3}\Delta^{1/2} \right),\\ \nonumber
\lambda_{2}&=\rho \lambda_{1}, \; \lambda_{3}=\rho^{2} \lambda_{1},
&\lambda_{5}&=\rho \lambda_{4},\; \lambda_{6}=\rho^{2} \lambda_{4},\qquad
\end{align*}
where $\Delta= 27(12\beta^2+ \gamma^2) $. Indeed, letting
$\alpha\to 0$ in  (\ref{bptsexp}), we see that the branchpoints
$\BH_{i}$ collide pairwise to give the $\lambda_{i}$ (see Figure
\ref{alphato0}). In Figure \ref{bpts0} we also give the monodromy
which is $[1,2,3]$, the same for every branchpoint; this can be
seen from an explicit calculation, but also by taking the limit of
the monodromies of $\hat{\mathcal{C}}$.

\subsection{The quotient with respect to $\texttt{C}_{3}$}\label{sec:quotient}

\begin{figure}
\begin{center}
 \hspace*{-6mm}
\includegraphics[height=250pt]{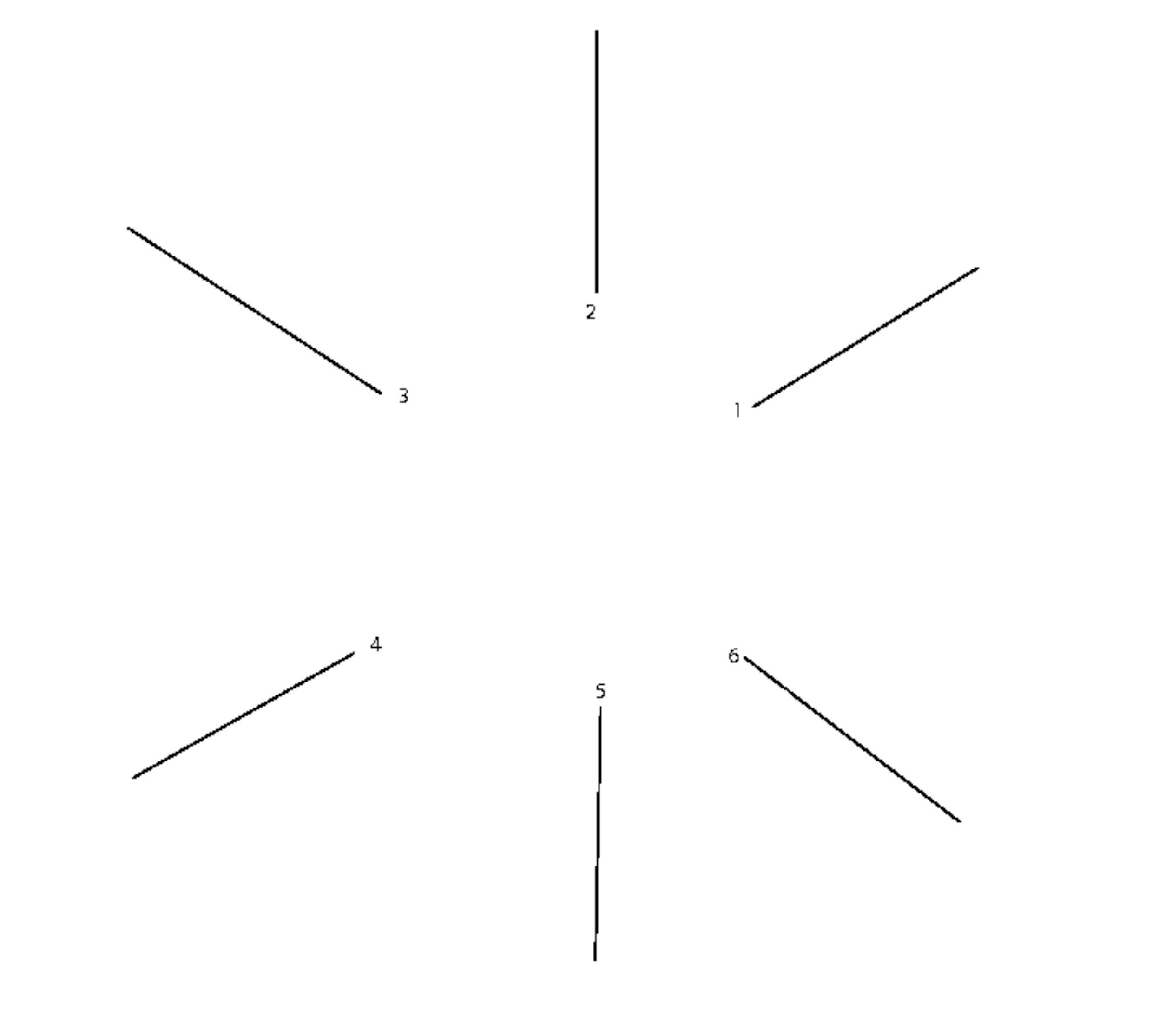}
\hspace*{-5mm}
\caption{Branchpoints and monodromy for $\mathcal{C}$}\label{bpts2}
\end{center}\end{figure}

We may form the quotient curve
$\mathcal{C}=\hat{\mathcal{C}}/\sigma$ with the covering map
\begin{equation}\label{quotient}
 \pi:\hat{\mathcal{C}}\longrightarrow \mathcal{C}:\qquad
    (\zeta,\eta)\longrightarrow (x,y)=\left(\frac{\eta}{\zeta},\;
    \beta(\zeta^{3}+\frac{1}{\zeta^{3}})\right).
\end{equation}
and the curve $\mathcal{C}$ is given by (\ref{quotientcurve}).
It is a {genus 2} (hence hyperelliptic) Riemann surface and $\pi$
is an unbranched covering. Viewing $\mathcal{C}$ as a 2-sheeted
cover of the Riemann sphere, it has six branchpoints whose
$x$-coordinates are
\begin{align}
\begin{split}\label{bpquotient}
B^{x}_{1}&=\frac{1}{6}\,\rho\,\delta_{-}^{\frac{1}{3}}-\frac {2\rho^{2}a}{\delta_{-}^{\frac{2}{3}}}, \qquad
B^{x}_{2}=\frac{1}{6}\,\rho\,\delta_{+}^{\frac{1}{3}}-\frac {2\rho^{2}a}{\delta_{+}^{\frac{2}{3}}}, \qquad
B^{x}_{3}=\frac{1}{6}\,\delta_{-}^{\frac{1}{3}}-\frac {2a}{\delta_{-}^{\frac{2}{3}}},  \\
B_{4}^{x}&=\frac{1}{6}\,\delta_{+}^{\frac{1}{3}}-\frac {2\,a}{\delta_{+}^{\frac{2}{3}}},  \qquad \quad\;\,
B_{5}^{x}=\frac{1}{6}\,\rho^{2}\,\delta_{-}^{\frac{1}{3}}-\frac {2\rho\, a}{\delta_{-}^{\frac{2}{3}}}, \qquad
B_{6}^{x}=\frac{1}{6}\,\rho^{2}\,\delta_{+}^{\frac{1}{3}}-\frac {2\rho\,a}{\delta_{+}^{\frac{2}{3}}},
\end{split}
\end{align}
where
\begin{equation*}
\delta_{\pm}= -108\,\gamma-216\beta\,i+12\,\sqrt {12\,\alpha^{3}+81\, \left( \gamma\pm2\beta\,i \right) ^{2}}.
\end{equation*}
As $\delta_{-}=\overline{\delta_{+}}$, these branchpoints can be
split into complex conjugate pairs
\begin{align}\label{bpquotientcc}
B^{x}_{6}&=\overline{B^{x}_{1}}, \qquad B^{x}_{5}=\overline{B^{x}_{2}} ,
\qquad B^{x}_{4}=\overline{B^{x}_{3}}.
\end{align}
The branchpoints $B_{i}$ \emph{are not} the images of the
branchpoints of $\hat{\mathcal{C}}$ under $\pi$. Figure
\ref{bpts2} shows again a qualitative sketch of the branchpoints
for the curve $\mathcal{}$ and their monodromy, with the same
choice of parameters of Figure \ref{bpts4}. Because the branch
points are not images of those of $\hat{\mathcal{C}}$ we observe
that there is little difference in the quotient curves for
$\alpha$ equalling or differing from zero.

A standard basis for the holomorphic differentials on
$\mathcal{C}$ is given by
\begin{equation}
{\mathbf{u}}_{1}=\frac{\de x}{y}, \quad
{\mathbf{u}}_{2}=\frac{x\;\de x}{y}.
\end{equation}
The differentials ${\hat{\mathbf{u}}_{2}}$ and
${\hat{\mathbf{u}}_{4}}$ on $\hat{\mathcal{C}}$ are invariant
under $\sigma$ and hence descend to differentials on
$\mathcal{C}$. In fact,
\begin{equation}\label{diffquots}
 \pi^{*}\;{\mathbf{u}}_{1}=-3{\hat{\mathbf{u}}_{2}}, \qquad \pi^{*}\;{\mathbf{u}}_{2}=-3
 {\hat{\mathbf{u}}_{4}}.
\end{equation}
This observation allows us to considerably simplify some
integrals, and hence the period matrix.

\subsection{Homology bases}\label{hombases} 
\begin{figure}
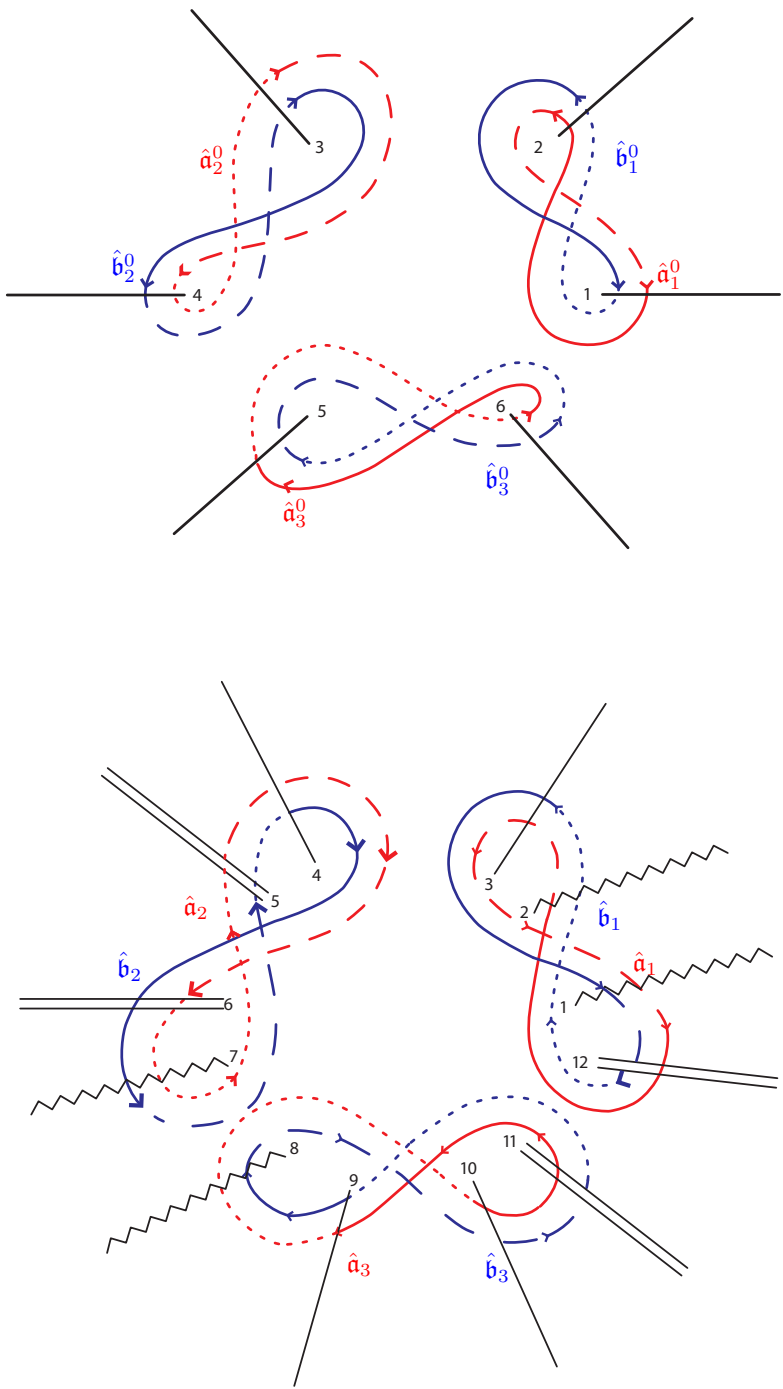

\hspace{-4cm}
{\begin{lpic}[draft,clean]{cycles0_123(12cm,)}
   \lbl[t]{135,75;$\color{red}{\azero_1}$}
   \lbl[t]{127,98;${\cblu{\bzero_1}}$}
   \lbl[t]{47,97;$\cred{\azero_2}$}
   \lbl[t]{30,77;${\cblu{\bzero_2}}$}
   \lbl[t]{63,29;$\color{red}{\azero_3}$}
   \lbl[t]{102,36;${\cblu{\bzero_3}}$}
   \end{lpic}}
{ \begin{lpic}[draft,clean]{cycles_123bis(12cm,)}
  \lbl[t]{131,98;$\color{red}{\ah_1}$}
   \lbl[t]{124,108;${\cblu{\bh_1}}$}
   \lbl[t]{43,109;$\cred{\ah_2}$}
   \lbl[t]{30,98;${\cblu{\bh_2}}$}
   \lbl[t]{75,39;$\color{red}{\ah_3}$}
   \lbl[t]{102,39;${\cblu{\bh_3}}$}   \end{lpic}}
\caption{Cyclic homology basis}\label{cyc123}
\end{figure}

By choosing an homology basis well
adapted to the symmetry at hand we may simplify many things. The
aim of this subsection is to construct one such basis. For the
case of our unbranched cover $\pi:\hat{\mathcal{C}}\rightarrow
\mathcal{C}$ it is known \cite{fay73} that there exists a basis
 $\{\hat{\mathfrak{a}}_0,\hat {\mathfrak{b}}_0 ,
 \hat{\mathfrak{a}}_1,\hat
 {\mathfrak{b}}_1,{\mathfrak{a}}_{2} ,\hat {\mathfrak{b}}_{2} ,\hat
{\mathfrak{a}}_{3},\hat {\mathfrak{b}}_{3} \}$ of
homology cycles for $\hat{\mathcal{C}}$ and
$\{{\mathfrak{a}}_0,{\mathfrak{b}}_0 ,{\mathfrak{a}}_1 ,
{\mathfrak{b}}_1,\}$ for $\mathcal{C}$ such that (for $1\le
j\le 3,\ 0\le s< 3$)
\begin{align}\label{fayfinal}
\sigma\sp{s}(\hat {\mathfrak{a}}_0)&\sim {\hat
{\mathfrak{a}}_0},&\sigma\sp{s}(\hat
{\mathfrak{a}}_j)&=\hat
{\mathfrak{a}}_{j+s},&
\sigma\sp{s}(\hat  {\mathfrak{b}}_0)&= {\hat
{\mathfrak{b}}_0},&\sigma\sp{s}(\hat
{\mathfrak{b}}_j)&=\hat {\mathfrak{b}}_{j+s} ,&
\\ \label{fayproj}
\pi(\hat
{\mathfrak{a}}_0)&={\mathfrak{a}}_0 ,&\pi(\hat
{\mathfrak{a}}_{j+s})&={\mathfrak{a}}_j ,&
\pi(\hat {\mathfrak{b}}_0)&=3\,
{\mathfrak{b}}_0,&\pi(\hat
{\mathfrak{b}}_{j+s})&={\mathfrak{b}}_j .&
\end{align}
Here $\sigma\sp{s}(\hat {\mathfrak{a}}_0)$ is homologous to
$\hat {\mathfrak{a}}_0$ and indices are understood to be
modulo 3. These requirements  do not uniquely determine a homology
basis and we may impose some extra conditions on this basis. The
condition we choose (and will see is possible) is that the limit
$\alpha\to 0$ of some cycles are mapped to some of the homology
basis in \cite{bren06}. This will enable us to relate the present
work with \cite{bren06} whose results we generalise. We also
observe that given two cycles $\hat{\mathfrak{a}}_{1}$,
$\hat{\mathfrak{b}}_{1}$ with canonical pairing then we may
simply define $\hat{\mathfrak{a}}_{2,3}$ via $\hat
{\mathfrak{a}}_{1+s}=\sigma\sp{s}(\hat
{\mathfrak{a}}_1)$ (and similarly for $
{\mathfrak{b}}_{2,3}$). Thus we seek two cycles
$\hat{\mathfrak{a}}_{1}$, $\hat{\mathfrak{b}}_{1} $
such that
$$\hat{\mathfrak{a}}_{1}\sp{0}:=\lim_{\alpha\rightarrow0}\hat{\mathfrak{a}}_{1}=
{\bf{a}}_{1}\sp{\textrm{BE06}},\qquad
\hat{\mathfrak{b}}_{1}\sp{0}:=
\lim_{\alpha\rightarrow0}\hat{\mathfrak{b}}_{1}=
{\bf{b}}_{1}\sp{\textrm{BE06}}.$$ Such cycles and their
corresponding images under $\sigma$ are shown in Figure
\ref{cyc123} alongside the $\alpha=0$ limit of these. One further
finds that we may choose $\hat{\mathfrak{a}}_{0}$ such that
$$\hat{\mathfrak{a}}_{0}\sp{0}:=\lim_{\alpha\rightarrow0}\hat{\mathfrak{a}}_{0}=
{\bf{a}}_{4}\sp{\textrm{BE06}}$$ and (\ref{fayfinal}) is satisfied.

It remains to find the cycle $\hat{\mathfrak{b}}_{0}$ and
this is the most difficult. From (\ref{fayfinal}) we see we wish a
cycle invariant under $\sigma$ and having canonical intersections
with the other cycles. Such is shown in Figure \ref{cyc0}
alongside $\hat{\mathfrak{a}}_{0}$ and their $\alpha=0$
limits. We record these results as
\begin{theorem} The homology cycles given in figures \ref{cyc123}
and \ref{cyc0} are canonical, satisfy (\ref{fayfinal}), and have
smooth limit $\alpha\rightarrow0$.
\end{theorem}

\begin{figure}
\begin{center}
   \begin{lpic}[draft,clean]{cycles0_0(10cm,)}
   \lbl[t]{107,100;$\color{red}{\azero_0}$}
   \lbl[t]{125,100;${\cblu{\bzero_0}}$}
   \end{lpic}
\vspace{-1cm}
   \begin{lpic}[draft,clean]{cycles_0(10cm,)}
      \lbl[t]{117,100;$\color{red}{\ah_0}$}
   \lbl[t]{134,104;${\cblu{\bh_0}}$}
   \end{lpic}
\caption{Cyclic homology basis}\label{cyc0}
\end{center}
\end{figure}

These cycles can be expanded in terms of ``basic arcs'' as
follows. Denote the arc between the branchpoints $\widehat{B}_{i}$
and $\widehat{B}_{j}$ on sheet $k$ by
\begin{equation*}
\gamma_k(i,j)=\mathrm{arc}_k ( \widehat{B}_{i},\widehat{B}_{j}  ),\quad i\neq j=1,\ldots,12.
\end{equation*}
Then we have the following
\begin{align}\label{arcexpsymmbasis}\begin{split}
\hat{{\mathfrak{a}}}_1&=\gamma_1(1,2)+\gamma_2(2,1),  \quad\;\quad
\hat{{\mathfrak{b}}}_1=\gamma_1(3,1)+\gamma_2(1,12)+\gamma_3(12,3),  \\
\hat{{\mathfrak{a}}}_2&= \gamma_2(5,6)+\gamma_3(6,5), \quad\quad\;
\hat{{\mathfrak{b}}}_2=  \gamma_2(7,5)+\gamma_3(5,4)+\gamma_1(4,7),    \\
\hat{{\mathfrak{a}}}_3&= \gamma_3(9,10)+\gamma_1(10,9),   \quad\;
\hat{{\mathfrak{b}}}_3=  \gamma_3(11,9)+\gamma_1(9,8)+\gamma_2(8,11),
\\
\hat{{\mathfrak{a}}}_0&=\gamma_1(3,10)+\gamma_3(10,9)+\gamma_1(9,8)
+\gamma_2(8,12)+\gamma_3(12,3)
, \\
\hat{{\mathfrak{b}}}_0&= \gamma_1(2,8)+\gamma_{2}(8,11)+\gamma_{3}(11,4)
+\gamma_{1}(4,7)+
\gamma_2(7,12)+\gamma_3(12,2)
.
\end{split}
\end{align}
and, as $\alpha\to 0$,
\begin{align}\begin{split}
\hat{{\mathfrak{a}}}_1\sp{0}&=\gamma_1(1,2)+\gamma_{2}(2,1),\quad
\hat{{\mathfrak{b}}}_1\sp{0}=\gamma_{1}(2,1)+\gamma_3(1,2),\\
\hat{{\mathfrak{a}}}_2\sp{0}&=\gamma_{2}(3,4)+\gamma_3(4,3),\quad
\hat{{\mathfrak{b}}}_2\sp{0}=\gamma_2(4,3)+\gamma_1(3,4),\\
\hat{{\mathfrak{a}}}_3\sp{0}&=\gamma_{3}(5,6)+\gamma_1(6,5),\quad
\hat{{\mathfrak{b}}}_3\sp{0}=\gamma_3(6,5)+\gamma_2(5,6), \label{arcexpzero} \\
\hat{{\mathfrak{a}}}_0\sp{0}&=\gamma_{3}(1,2)+\gamma_1(2,6)+\gamma_{3}(6,5)
+\gamma_2(5,1),\\
\hat{{\mathfrak{b}}}_0\sp{0}&=\gamma_{3}(1,2)+\gamma_{1}(2,5)+\gamma_2(5,6)
+\gamma_3(6,3)+\gamma_1(3,4)+\gamma_2(4,1).
\end{split}
\end{align}

\begin{figure}
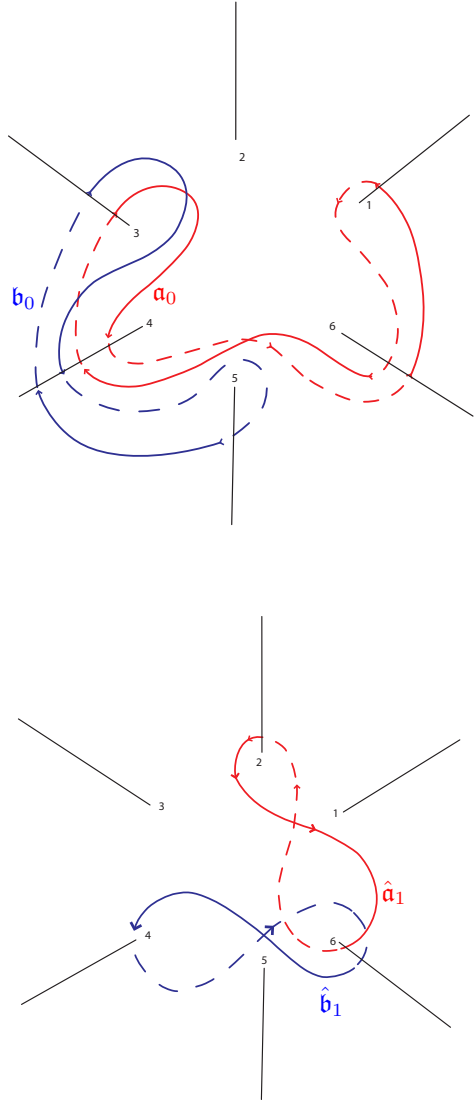

\begin{center}
\begin{lpic}[draft,clean]{cycles_g2_0(8cm,)}
    \lbl[t]{63,90;$\cred{\aq_0}$}
   \lbl[t]{22,90;${\cblu{\bq_0}}$}
   \end{lpic}
   \begin{lpic}[draft,clean]{cycles_g2_1(8cm,)}
  \lbl[t]{134,78;$\color{red}{\ah_1}$}
      \lbl[t]{115,46;${\cblu{\bh_1}}$}   \end{lpic}
\caption{Cyclic homology basis for the quotient curve $\mathcal{C}$}\label{cycg2}
\end{center}
\end{figure}

We may complete the specification (\ref{fayproj}) of homology
bases by projecting the cycles of Figures \ref{cyc123},
\ref{cyc0}. The fact that the branchpoints of $\hat{\mathcal{C}}$
do not get mapped by $\pi$ to branchpoints of $\mathcal{C}$ makes
the projection less straightforward\footnote{This has been
implemented in Maple.}. The results are shown in Figure
\ref{cycg2}. We therefore have a homology basis for the
hyperelliptic curve $\mathcal{C}$ differing from standard ones. As
we shall see however, the bases chosen allow us to simply relate
the period matrices and other quantities of $\hat{\mathcal{C}}$
and $\mathcal{C}$. With the same notation as above, the arc
expansion for these cycles is then
\begin{align}\label{basisqexp} \begin{split}
{{\mathfrak{a}}}_1&=\gamma_{1}(2,6)+\gamma_{2}(6,2),\\
{{\mathfrak{b}}}_1&=\gamma_{1}(6,4)+\gamma_{2}(4,6),\\
{{\mathfrak{a}}}_0&=\gamma_{1}(3,4)+\gamma_{2}(4,6)+\gamma_{1}(6,1)
+\gamma_{2}(1,6)+\gamma_{1}(6,4)+\gamma_{2}(4,3),\\
{{\mathfrak{b}}}_0&=\gamma_{1}(3,4)+\gamma_{2}(4,5)+\gamma_{1}(5,4)
+\gamma_{2}(4,3).
\end{split}\end{align}

\subsection{Period matrices}
We shall now relate the period matrices of $\hat{\mathcal{C}}$ and
$\mathcal{C}$ and then use the symmetries of $\hat{\mathcal{C}}$
to further restrict the periods involved. If
$\{\hat{\mathfrak{a}}_\mu,\hat{\mathfrak{b}}_\mu \}$
are the canonical homology basis introduced earlier and
$\{\hat{\mathbf{u}}_j\}$ any basis of holomorphic differentials
for our Riemann surface $\hat{\mathcal{C}}$ we have the matrix of
periods
\begin{equation}\begin{pmatrix}\oint_{\hat{\mathfrak{a}}_\mu}\hat{\mathbf{u}}_j\\
\oint_{\hat{\mathfrak{b}}_\mu}\hat{\mathbf{u}}_j\end{pmatrix}=
\begin{pmatrix}\hat{\mathcal{A}}\\
\hat{\mathcal{B}}\end{pmatrix}=\begin{pmatrix}1\\
\hat{\tau}\end{pmatrix}\hat{\mathcal{A}}
\end{equation}
with $\hat{\tau}=\hat{\mathcal{B}}\hat{\mathcal{A}}\sp{-1}$
the period matrix. The period matrix and periods
$\hat{\mathcal{A}}$, $\hat{\mathcal{B}}$ are our focus here.

To understand the connection between the period matrices
$\hat{\tau}$ of $\hat{\mathcal{C}}$ and ${\tau} $ of
$\mathcal{C}$ we may first focus on the
$\hat{\mathfrak{a}}$-normalized differentials. If $\hat
{\mathbf{v}}_i$ are the $\hat{\mathfrak{a}}$-normalized
differentials for $\hat{\mathcal{C}}$, then
$$\delta_{i,j+s}=\int_{\hat {\mathfrak{a}}_{j+s}}\hat {\mathbf{v}}_i=
\int_{\sigma\sp{s}(\hat {\mathfrak{a}}_j)}\hat {\mathbf{v}}_i=\int_{\hat
{\mathfrak{a}}_j}(\sigma\sp{s})\sp{\ast}\hat {\mathbf{v}}_i=\int_{\hat
{\mathfrak{a}}_j}\hat {\mathbf{v}}_{i-s},$$ and we find that
\begin{equation}(\sigma\sp{s})\sp{\ast}\hat {\mathbf{v}}_0= \hat
{\mathbf{v}}_0,\qquad(\sigma\sp{s})\sp{\ast}\hat {\mathbf{v}}_i=\hat {\mathbf{v}}_{i-s}.
\label{gpdiff} \end{equation} If ${\mathbf{v}}_i$ are the normalized
differentials for $\mathcal{C}$, then
$$
\delta_{ij}=\int_{{\mathfrak{a}}_j}{\mathbf{v}}_i=\int_{\pi(\hat
{\mathfrak{a}}_{j+s})}{\mathbf{v}}_i
=\int_{{\mathfrak{a}}_{j+s}}\pi\sp\ast ({\mathbf{v}}_i)$$ shows that
$$\pi\sp\ast ({\mathbf{v}}_i)=\hat {\mathbf{v}}_i+(\sigma)\sp{\ast}\hat {\mathbf{v}}_i+(\sigma\sp{2})\sp{\ast}\hat {\mathbf{v}}_i$$
 and similarly that
$$\pi\sp\ast ({\mathbf{v}}_0)=\hat {\mathbf{v}}_0.$$

From (\ref{gpdiff}) we have an action of $\texttt{C}_3$ on
$\text{Jac}(\hat{\mathcal{C}})$ which lifts to an automorphism of
${\mathbb C}\sp{ 4}$ by
\begin{equation}\label{symmjac}
\sigma(\hat z)=\sigma(\hat z_0,\hat z_{1},\hat
z_{2},\hat z_{3})=(\hat z_0,\hat z_{3},\hat
z_{2},\hat z_{1})
\end{equation}
With the choices above (things are different for
$\hat{\mathfrak{b}}$-normalization) we may lift the map
$\pi\sp\ast: \Jac(\mathcal{C})\rightarrow \Jac(\hat{\mathcal{C}})$
to $\pi\sp\ast:{\mathbb C}\sp{2}\rightarrow {\mathbb C}\sp{ 4}$,
$$
\pi\sp\ast(z)= \pi\sp\ast(z_0,z_1)= (3\,z_0,z_1,z_1,z_{1})=\hat
z.$$ If we denote points of the Jacobian in characteristic
notation by
$$\left[ \begin{matrix}\alpha \\ \beta
\end{matrix}\right]_{\tau}=\alpha{\tau } +\beta$$
($\alpha$,  $\beta\in{\mathbb R}\sp{ 2}$) then
$$\pi\sp\ast\left[ \begin{matrix}\alpha_0&\alpha_1 \\
\beta_0&\beta_1
\end{matrix}\right]_{\tau}=\left[ \begin{matrix}\alpha_0&\alpha_1&\alpha_1&\alpha_1
 \\3\, \beta_0&\beta_1&\beta_1&\beta_1
\end{matrix}\right]_{\hat\tau}.$$
The period matrices for the two curves are related by \cite{fay73}
\begin{equation}\label{cyclicpm}
\hat{\tau}
=\left( \begin{array}{cccc} a&b&b&b\\
b&c&d&d\\
b&d&c&d\\
b&d&d&c
\end{array}\right),\qquad
\tau
=\left( \begin{array}{cc} \frac13 a&b\\
b&c+2d
\end{array}\right) .\end{equation}
The point to note is that although the period matrix for
$\hat{\mathcal{C}}$ involves integrations of differentials that do
not reduce to hyperelliptic integrals, the combination of terms
appearing in the reduction can be expressed in terms of
hyperelliptic integrals. This is a definite simplification.
Further the $\Theta$ function defined by $\hat\tau$ has the
symmetries
$$\Theta(\hat z|\hat\tau)=\Theta(\sigma\sp{s}(\hat z)|\hat\tau )$$
for all $\hat z\in{\mathbb C}\sp{4}$. In particular, the $\Theta$
divisor is fixed under $\texttt{C}_3$.

Now we turn to the symmetries of $\hat{\mathcal{C}}$ to simplify
the calculation of periods. If $\psi$ is any automorphism of
$\hat{\mathcal{C}}$ then $\psi$ acts on
$H_1(\hat{\mathcal{C}},\mathbb{Z})$ and the holomorphic
differentials by
$$\psi_\ast\begin{pmatrix}\hat{\mathfrak{a}}_\mu\\
\hat{\mathfrak{b}}_\mu\end{pmatrix}=\begin{pmatrix}{A}&B\\
C&D\end{pmatrix}\begin{pmatrix}\hat{\mathfrak{a}}_\mu\\
\hat{\mathfrak{b}}_\mu\end{pmatrix},\qquad
\psi\sp\ast\hat{\mathbf{u}}_j=\hat{\mathbf{u}}_k
L\sp{k}_j,$$ where $\begin{pmatrix}{A}&B\\
C&D\end{pmatrix}\in Sp(8,\mathbb{Z})$ and $L\in GL(4,\mathbb{C})$.
Then from
\begin{equation*}\oint_{\psi_\ast\gamma}\hat{\mathbf{u}}=\oint_\gamma\psi\sp\ast\hat{\mathbf{u}}
\end{equation*}
we obtain
\begin{align}
\begin{pmatrix}{A}&B\\
C&D\end{pmatrix}\begin{pmatrix}\hat{\mathcal{A}}\\
\hat{\mathcal{B}}\end{pmatrix}&=\begin{pmatrix}\hat{\mathcal{A}}\\
\hat{\mathcal{B}}\end{pmatrix}L .
\end{align}
Thus, for example, from the definition of our homology basis,
$$
M_{\sigma}:=
\left(
\begin{array}{cccccccc}
 1 &  0 & 0 & 0  & 0 & 0  & 0 &0  \\
 0 & 0  & 1 & 0 & 0  & 0 & 0  & 0   \\
 0 & 0  & 0 & 1 & 0  & 0 & 0  & 0   \\
 0 & 1 & 0 & 0 & 0  & 0 & 0  & 0   \\
 0 & 0  & 0 & 0 & 1  & 0 & 0  & 0   \\
 0 & 0  & 0 & 0 & 0  & 0 & 1  & 0   \\
 0 & 0  & 0 & 0 & 0  & 0 & 0  & 1   \\
 0 & 0  & 0 & 0 & 0  & 1 & 0  & 0
\end{array}
\right),
$$
while $\sigma$ acts on the differentials (\ref{diffxhat}) as
follows
\begin{align}
\sigma^{*}{\hat{\mathbf{u}}_{1}}&=\rho^{2}{\hat{\mathbf{u}}_{1}},
& \sigma^{*}{\hat{\mathbf{u}}_{2}}&={\hat{\mathbf{u}}_{2}}, &
\sigma^{*}{\hat{\mathbf{u}}_{3}}&=\rho\;{\hat{\mathbf{u}}_{3}},
&\sigma^{*}{\hat{\mathbf{u}}_{4}}&={\hat{\mathbf{u}}_{4}}.
\end{align}
Let us denote the $\hat{\mathfrak{a}}_i$ integrals of
${\hat{\mathbf{u}}_{1}}$ by $z_i$ and the corresponding
$\hat{\mathfrak{b}}_i$ integrals by $Z_i$; and similarly those of
${\hat{\mathbf{u}}_{2}}$, ${\hat{\mathbf{u}}_{3}}$ and
${\hat{\mathbf{u}}_{4}}$ by $x_i$, $X_i$, $w_i$, $W_i$ and $y_i$,
$Y_i$ respectively. Then $\hat{\mathcal{A}}=(\mathbf{z},
\mathbf{x}, \mathbf{w}, \mathbf{y})$ (and analogously for
$\hat{\mathcal{B}}$). A symmetry relates these various periods.
The symmetry $\sigma$ restricts the matrices of periods to take
the following form
\begin{equation}\label{pihatsigma}
\hat{\mathcal{A}}=\left( \begin{array}{cccc}
0&x_{{0}}&0&y_{{0}}\\
z_{{1}}&x_{{1}}&w_{{1}}&y_{{1}}\\
{\rho}^{2}z_{{1}}&x_{
{1}}&\rho\,w_{{1}}&y_{{1}}\\
\rho\,z_{{1}}&x_{{1}}&{\rho}^{2}w_{{1}}&y_{{1}}\end {array}\right), \quad
\hat{\mathcal{B}}= \left(\begin{array}{cccc}
0&X_{{0}}&0&Y_{{0}}\\
Z_{{1}}&X_{{1}}&W_{{1}}&Y_{{1}}\\
{\rho}^{2}Z_{{1}}&X_{{1}}&\rho\,W_{{1}}&Y_{{1}}\\
\rho\,Z_{{1}}&X_{{1}}&{\rho}^{2}W_{{1}}&Y_{{1}}
\end {array} \right).
\end{equation}
For instance, we get
$$z_{0}=\oint_{\hat {\mathfrak{a}}_{0}}{\hat{\mathbf{u}}_{1}}
=\oint_{\sigma( {\mathfrak{a}}_{0})}{\hat{\mathbf{u}}_{1}}
=\oint_{\hat {\mathfrak{a}}_{0}}\sigma\sp\ast
{\hat{\mathbf{u}}_{1}} =\oint_{\hat
{\mathfrak{a}}_{0}}\rho\sp2  {\hat{\mathbf{u}}_{1}}
=\rho\sp2 z_0,$$ and so $z_{0}=0$, while
$$z_2=\oint_{\hat {\mathfrak{a}}_{2}}{\hat{\mathbf{u}}_{1}}
=\oint_{\sigma({\mathfrak{a}}_{0})}{\hat{\mathbf{u}}_{1}}=
\oint_{\hat {\mathfrak{a}}_{0}}\sigma\sp\ast{\hat{\mathbf{u}}_{1}}
=\oint_{\hat {\mathfrak{a}}_{0}}\rho\sp2  {\hat{\mathbf{u}}_{1}}
=\rho\sp2\,z_1,
$$
so leading to \eqref{pihatsigma}.

We now use the other symmetries to further restrict the matrices
of periods. The chief difficulty in this approach is in
calculating the actions on the homology. In the present setting we
find that
\begin{equation}\label{tauM}
M_{\varphi}:=\left( \begin{array}{rrrrrrrr}
-1&0&0 &0&0&0&0&0\\
0&0&-1&0&0&0&0&0\\
0&-1&0 &0&0&0&0&0\\
0&0&0&-1&0&0&0&0\\
0&0&0&0&-1&0&0&0\\
0&0&0&0&0&0&-1&0\\
0&0&0&0&0&-1&0&0\\
0&0&0&0&0&0&0&-1
\end {array} \right),\
M_{\tau}:=\left( \begin {array}{rrrrrrrrr}
2&0&0&0&0&-1&-1&-1\\
1&0&1&1&-1&0&1&1\\
1&1&0&1&-1&1&0&1\\
1&1&1&0&-1&1&1&0\\
6&1&1&1&-2&-1&-1&-1\\
1&0&0&0&0&0&-1&-1\\
1&0&0&0&0&-1&0&-1\\
1&0&0&0&0&-1&-1&0
\end {array}
 \right),
\end{equation}
with
\begin{align}\label{phi13}
\varphi^{*}\hat{\mathbf{u}}_{1}&=\hat{\mathbf{u}}_{3},
&\varphi^{*}\hat{\mathbf{u}}_{2}&=-\hat{\mathbf{u}}_{2},
&\varphi^{*}\hat{\mathbf{u}}_{3}&=\hat{\mathbf{u}}_{1},
&\varphi^{*}\hat{\mathbf{u}}_{4}&=-\hat{\mathbf{u}}_{4},\\
\tau^{*}\hat{\mathbf{u}}_{1}&=\overline{\hat{\mathbf{u}}_{3}},
&\tau^{*}\hat{\mathbf{u}}_{2}&=-\overline{\hat{\mathbf{u}}_{2}},
&\tau^{*}\hat{\mathbf{u}}_{3}&=\overline{\hat{\mathbf{u}}_{1}},
&\tau^{*}\hat{\mathbf{u}}_{4}&=-\overline{\hat{\mathbf{u}}_{4}}.
\end{align}
The $\varphi$ symmetry simplifies the matrices of periods to the form
\begin{equation*}
\hat{\mathcal{A}}=\left( \begin{array}{cccc}
0&x_{{0}}&0&y_{{0}}\\
z_{{1}} & x_{{1}}& -\rho^{2}z_{1} & y_{{1}}\\
{\rho}^{2}z_{{1}}&x_{{1}}&-z_{1}&y_{{1}}\\
\rho\,z_{{1}}&x_{{1}}&-\rho z_{1}&y_{{1}}\end {array}\right),\quad
\hat{\mathcal{B}}= \left(\begin{array}{cccc}
0&X_{{0}}&0&Y_{{0}}\\
Z_{{1}}&X_{{1}}&-\rho^{2}Z_{1}&Y_{{1}}\\
{\rho}^{2}Z_{{1}}&X_{{1}}&-Z_{1}&Y_{{1}}\\
\rho\,Z_{{1}}&X_{{1}}&-\rho Z_{1}&Y_{{1}}
\end {array} \right)
\end{equation*}
while the real involution relates the entries via
\begin{align}
Z_{1}&=-(\bar{w}_{1}+z_{1})=\rho\bar{z}_{1}-z_{1}, &W_{1}&=-(\bar{z}_{1}+w_{1})=\rho^{2}z_{1}-\bar{z}_{1}, \label{ZWtau}\\
3X_{1}&=2x_{0}+\bar{x}_{0},  & 3Y_{1}&=2y_{0}+\bar{y}_{0}, \\
X_{0}&=2x_{1}+x_{0}+\bar{x}_{1}+2X_{1}, &
Y_{0}&=2y_{1}+y_{0}+\bar{y}_{1}+2Y_{1}.
\end{align}
Thus all the periods are determined in terms of $z_1$, $x_1$, $x_0$, $y_1$ and $y_0$. Finally there is
the bilinear relation
$$0=x_0Y_0-y_0X_0+3(x_1Y_1-y_1X_1).$$
Calculating the period matrix
$\hat{\tau}=\hat{\mathcal{B}}\hat{\mathcal{A}}\sp{-1}$ we
obtain the form (\ref{cyclicpm}) with
\begin{align}%
a&=\dfrac{x_1Y_0-X_0y_1}{x_1y_0-x_0y_1}, &\quad b&=\dfrac{x_1Y_1-X_1y_1}{x_1y_0-x_0y_1},
\label{abcd} \\
c&=\dfrac{2}{3}\frac{Z_1}{z_1}-\frac13\frac{x_0Y_1-X_1y_0}{x_1y_0-x_0y_1},
&\quad d&=-\dfrac{1}{3}\frac{Z_1}{z_1}-\frac13\frac{x_0Y_1-X_1y_0}{x_1y_0-x_0y_1}.
\nonumber
\end{align}

\subsubsection{Weierstrass-Poincar\'e reduction}
We remark in passing that the symplectic matrix
\begin{equation}
T=\left(  \begin{array}{cccccccc}
0&  1& 1& 1& 0& 0& 0& 0  \\
1&  0& 0& 0& 0& 0& 0& 0  \\
0&  0& 0& 0& 0&-2& 1& 1  \\
0&  0& 1&-1& 0& 0& 0& 0  \\
0&  0& 0& 0& 0& 1& 0& 0  \\
0&  0& 0& 0& 1& 0& 0& 0  \\
0&  0& 0&-1& 0& 0& 0& 0  \\
0&  0& 0& 0& 0&-1& 1& 0    \end{array} \right)=\left( \begin{array}{cc} A&B\\C&D \end{array}\right)
\end{equation}
transforms the period matrix (\ref{cyclicpm}) as
$$\hat{\tau}\rightarrow (C+D\hat{\tau} )(A+B\hat{\tau} )^{-1}=
\left(   \begin{array}{cc} -{\tau}\sp{c\, -1}&Q\\
   Q^T&\mathfrak{T} \end{array} \right)$$
where $Q=\mathrm{Diag} (-1/3,0)$ and
$$\mathfrak{T}'=\left( \begin{array}{cc} \frac{c-d}{6}&\frac12\\\\\frac12&-\frac{1}{2(c-d)}
\end{array}\right).$$
From this we deduce that \begin{equation}
\Im(c-d)\ne0.\label{imcd}
\end{equation}

\subsubsection{The antiholomorphic involution for $\mathcal{C}$}
We have seen that the spectral curve $\hat{\mathcal{C}}$ has real
structure (\ref{realinv}). This real structure is inherited by
$\mathcal{C}$ where we have the antiholomorphic involution
\begin{equation}\label{realinvC}
\tau':\ (x,y)\mapsto (\bar x,-\bar y).
\end{equation}
The effect of this is on the homology above is to reflect in the
$x$-axis and change sheet. Specifically we find that
\begin{equation}\label{MtauC}
M_{\tau'}=\begin{pmatrix}
 2&0&0&-3\\ 1&2&-3&2
\\ 2&1&-2&-1\\ 1&0&0&-2
\end{pmatrix},\qquad M_{\tau'}\sp2=\rm{Id},\qquad M_{\tau'}JM_{\tau'}\sp{T}=-J,
\end{equation}
and where the last identity reflects that $M_{\tau'}$ is
antiholomorphic. On the holomorphic differentials we have the
simple action
\begin{equation*}
 \tau'^{*}\;{\mathbf{u}}_{1}=-{{\mathbf{u}}_{1}}, \qquad \tau'^{*}\;{\mathbf{u}}_{2}=-
 {{\mathbf{u}}_{2}}.
\end{equation*}

\subsection{The Fay-Accola theorem}
Having established homology bases (\ref{fayfinal}, \ref{fayproj})
and the relationship these entail for the corresponding period
matrices (\ref{cyclicpm}) of the curves,  we next recall the
striking theorem of Fay and Accola applied to our present setting.
\begin{theorem}[\bf Fay-Accola] \label{fayaccola}
With respect to the ordered canonical homology bases $\{
\hat{\mathfrak{a}}_i,\hat{\mathfrak{b}}_i \}$
constructed above and for arbitrary $\boldsymbol{ z}=\in
\mathbb{C}^2$ we have that
\begin{equation}
\frac{\theta[\hat e](\pi\sp\ast\boldsymbol{ z};\hat{\tau})}
{\prod_{k=0}^{2}\theta\left[\begin{matrix}0&0
\\ \frac{k}{3}&0 \end{matrix}\right]\left(\boldsymbol{ z};\tau
\right)}
=c_0(\widehat{\tau}
) \label{fafactora}
\end{equation}
is a non-zero modular constant
$c_0(\hat{\tau}
) $ independent of $\boldsymbol{ z}$. Here $\hat{\tau}$ and
${\tau}$ are  the $\mathfrak{a}$-normalized period matrices
for the respective curves given in the above bases and
\begin{equation*}
\hat e=\pi\sp\ast(e):=\pi^{*}\left ( \frac{3-1}{2\cdot3}, 0 \right) =
\left ( 1, 0, 0, 0\right)\equiv \mathbf{0}.
\end{equation*}
\end{theorem}
The significance of this theorem is that for flows on the Jacobian
of $\hat{\mathcal{C}}$ that arise as pullbacks of flows on the
Jacobian of ${\mathcal{C}}$ we may reduce the theta functions to
those of the hyperelliptic  spectral curve. We have stated in the
introduction that such a connection holds,
$$\lambda\widehat{\boldsymbol{U}}-
\widetilde{\boldsymbol{K}}=
\pi\sp\ast(\lambda \boldsymbol{U} -{\boldsymbol{K}}_{\infty_+}+e),
$$
and we now describe the quantities appearing in this.

\subsection{The vector of Riemann constants}
To construct the Baker-Akhiezer function for monopoles there is a
distinguished point
$\widetilde{\boldsymbol{K}}\in\Jac(\hat{\mathcal{C}})$ that
Hitchin uses to identify degree $\hat{g}-1$ line bundles with
$\Jac(\hat{\mathcal{C}})$. For $n\ge3$ this point is a singular
point of the theta divisor, $ \widetilde{\boldsymbol{K}}\in
\Theta_{\rm singular}$ \cite{bren06}. If we denote the Abel map by
$$\mathcal{A}_{\hat{Q}}(\hat{P})=\int_{\hat{Q}}\sp{\hat{P}}\hat{u}_i$$
then \begin{equation}\label{tildeK}\widetilde{\boldsymbol{K}}=
\boldsymbol{\hat{K}}_{\hat{Q}}+\mathcal{A}_{\hat{Q}}\left((n-2)
\sum_{k=1}\sp{n}\hat{\infty}_k\right).\end{equation} Here
$\boldsymbol{\hat{K}}_{\hat{Q}}$ is the vector of Riemann
constants for the curve $\hat{\mathcal{C}}$ and $\hat{\infty}_k$
are the points above infinity for the curve. If
$\mathcal{K}_{\hat{\mathcal{C}}}$ is the canonical divisor of the
curve then
$\mathcal{A}_{\hat{Q}}(\mathcal{K}_{\hat{\mathcal{C}}})=-2\boldsymbol{\hat{K}}_{\hat{Q}}$.
The righthand side of (\ref{tildeK}) is in fact independent of the
base point $\hat{Q}$ in its definition. Let $\pi(\hat{\infty}_k)=
\infty_+\in \mathcal{C}$ denote the projection of the points at
infinity. Then \cite{bra10} shows that
\begin{equation}\label{tildeKpb}\widetilde{\boldsymbol{K}}=
\pi\sp\ast({\boldsymbol{K}}_{\infty_+})-\hat e =
\pi\sp\ast({\boldsymbol{K}}_{\infty_+}-e),\end{equation} where the
half-period $\hat e$ has been identified in (\cite{fay73}). Thus
we need to calculate the vector of Riemann constants (for the
homology bases constructed) for the genus 2 curve
(\ref{quotientcurve}) and where the basepoint for the Abel map is
$\infty_{+}$.  It will be easier for our calculations to choose
one of the branchpoints, say $B_{1}$, and then to obtain the
vector of Riemann constants with respect to $\infty_{+}$ by using
the relation
\begin{equation}\label{b1infty}
\boldsymbol{K}_{\infty_{+}}=\abelmap_{\infty_{+}}(B_{1})+\boldsymbol{K}_{B_{1}},
\end{equation}
where $\abelmap_{\infty_{+}}$ is the Abel map with basepoint
${\infty_{+}}$.
\subsubsection{The vector $\boldsymbol{K}_{B_{1}}$}
We  begin by expressing the integrals over our homology cycles in
a simple form as integrals between branch points. Let
$\gamma_{i}(j,k)$ denote the path going from branchpoint $B_{j}$
to $B_{k}$ on the cut plane of Figure \ref{bpts2} corresponding to
sheet $i$. With this notation the cycles (\ref{basisqexp}) can be
expressed as
\begin{align*}
\aq_{1}&=\gamma_{1}(2,1)+\gamma_{1}(1,6)+\gamma_{2}(6,1)+\gamma_{2}(1,2),\\
\bq_{1}&=\gamma_{1}(6,5)+\gamma_{1}(5,4)+\gamma_{2}(4,5)+\gamma_{2}(5,6),\\
\aq_{0}&=\gamma_{1}(3,4)+\gamma_{2}(4,5)+\gamma_{2}(5,6)+\gamma_{1}(6,1)+
\gamma_{2}(1,6)+\gamma_{1}(6,5)+\gamma_{1}(5,4)+\gamma_{2}(4,3),\\
\bq_{0}&=\gamma_{1}(3,4)+\gamma_{2}(4,5)+\gamma_{1}(5,4)+\gamma_{2}(4,3).
\end{align*}
These expressions may be further simplified using the
hyperelliptic involution $J:(x,y)\rightarrow(x,-y)$, giving $
\gamma_{k+1}(2j,2j+1)=J\gamma_{k}(2j,2j+1)$. Also, as the sum of
the cycle encircling $B_1$ and $B_2$ together with the cycle
encircling $B_3$ and $B_4$  and the cycle encircling $B_5$ and
$B_6$ is homologically trivial, we find that
 relations
\begin{align*}
\int_{\gamma_{k}(6,5)}\omega&=\int_{\gamma_{k}(1,2)}\omega+
\int_{\gamma_{k}(3,4)}\omega,
\end{align*}
for any holomorphic differential $\omega$. Similar expressions
result from other homologically trivial choices of cycles.
These yield
\begin{align*}
\int_{\mathfrak{a}_{1}}\boldsymbol{\omega}&=
2\left(\int_{\gamma_{1}(2,1)}\boldsymbol{\omega}+\int_{\gamma_{1}(2,3)}\boldsymbol{\omega}+
\int_{\gamma_{1}(4,5)}\boldsymbol{\omega}\right),\\
\int_{\mathfrak{b}_{1}}\boldsymbol{\omega}&=
2\left(\int_{\gamma_{1}(1,2)}\boldsymbol{\omega}+\int_{\gamma_{1}(3,4)}\boldsymbol{\omega}+
\int_{\gamma_{1}(5,4)}\boldsymbol{\omega}\right),\\
\int_{\mathfrak{a}_{0}}\boldsymbol{\omega}&=
2\left(\int_{\gamma_{1}(3,4)}\boldsymbol{\omega}+\int_{\gamma_{1}(1,2)}\boldsymbol{\omega}+
\int_{\gamma_{1}(3,4)}\boldsymbol{\omega}+\int_{\gamma_{1}(5,4)}\boldsymbol{\omega}
\int_{\gamma_{1}(3,2)}\boldsymbol{\omega}+\int_{\gamma_{1}(5,4)}\boldsymbol{\omega}\right),\\
\int_{\mathfrak{b}_{0}}\boldsymbol{\omega}&=
2\left(\int_{\gamma_{1}(3,4)}\boldsymbol{\omega}+\int_{\gamma_{1}(5,4)}\boldsymbol{\omega}\right),
\end{align*}
and these may be inverted to obtain the integrals between
branchpoints.  If the $\boldsymbol{\omega}$ are taken to be
$\mathfrak{a}$-normalized differentials we have
\begin{align*}
\int_{\gamma_{1}(1,2)}\boldsymbol{\omega}&=\frac{1}{2}(\tau^{(0)}+\tau^{(1)}),
&\int_{\gamma_{1}(2,3)}\boldsymbol{\omega}&=\frac{1}{2}(e^{(0)}+\tau^{(0)}+\tau^{(1)}),
&\int_{\gamma_{1}(3,4)}\boldsymbol{\omega}&=\frac{1}{2}(e^{(0)}+e^{(1)}+\tau^{(0)}),\\
\int_{\gamma_{1}(4,5)}\boldsymbol{\omega}&=\frac{1}{2}(e^{(0)}+e^{(1)}),
&\int_{\gamma_{1}(5,6)}\boldsymbol{\omega}&=\frac{1}{2}(e^{(0)}+e^{(1)}+\tau^{(1)}),
\end{align*}
where $\tau^{(i)}$ and $e^{(i)}$ are the appropriate rows of the
period and identity matrices. Thus one can easily deduce the image
of each branchpoint under the Abel map (with basepoint $B_{1}$).
We obtain (together with their characteristic form)
\begin{align*}
\abelmap_{B_{1}}(B_{1})&=0
&\equiv&\ \ \left[ \begin{array}{cc}
0&0\\
0&0
\end{array}\right],\
&\abelmap_{B_{1}}(B_{2})=\frac{1}{2}(\tau^{(0)}+\tau^{(1)})
&\equiv&\frac{1}{2}\left[ \begin{array}{cc}
1&1\\
0&0
\end{array}\right],\\
\abelmap_{B_{1}}(B_{3})&=\frac{1}{2}e^{(0)}
&\equiv&\frac{1}{2}\left[ \begin{array}{cc}
0&0\\
1&0
\end{array}\right],\
&\abelmap_{B_{1}}(B_{4})=\frac{1}{2}(e^{(1)}+\tau^{(0)})
&\equiv&\frac{1}{2}\left[ \begin{array}{cc}
1&0\\
0&1
\end{array}\right],\\
\abelmap_{B_{1}}(B_{5})&=\frac{1}{2}(e^{(0)}+\tau^{(0)})
&\equiv&\frac{1}{2}\left[ \begin{array}{cc}
1&0\\
1&0
\end{array}\right],\
&\abelmap_{B_{1}}(B_{6})=\frac{1}{2}(e^{(1)}+\tau^{(0)}+\tau^{(1)})
&\equiv&\frac{1}{2}\left[ \begin{array}{cc}
1&1\\
0&1
\end{array}\right].
\end{align*}
Following an argument of Farkas and Kra (\cite{fk80} VII.1.2), the
vector of Riemann constants takes the form
\begin{align}
\boldsymbol{K}_{B_{1}}&=-(\abelmap_{B_{1}}(B_{5})+\abelmap_{B_{1}}(B_{6}))
=\frac{1}{2}(e^{(0)}+e^{(1)}+\tau^{(1)})\equiv\frac{1}{2}\left[ \begin{array}{cc}
0&1\\
1&1
\end{array}\right].
\end{align}
\subsubsection{The vector $\boldsymbol{K}_{\infty_{+}}$ }
Once we have calculated the vector of Riemann constants with $B_{1}$ as basepoint,
 we can change its basepoint making use of equation
 \eqref{b1infty}. One finds
\begin{equation}
\abelmap_{\infty_{+}}(B_{1})=\frac{2}{3}e^{(1)}+\frac{1}{2}\tau^{(1)}\equiv\left[ \begin{array}{cc}
\frac{1}{2}&0\\
\frac{2}{3}&0
\end{array}\right],
\end{equation}
and consequently
\begin{equation}\label{vrc}
{\boldsymbol{K}}_{\infty_+}=\frac{1}{6}e^{(1)}+\frac{1}{2}e^{(2)}+\frac{1}{2}\tau^{(1)}+
\frac{1}{2}\tau^{(2)}\equiv\left[ \begin{array}{cc}
\frac{1}{2}&\frac{1}{2}\\
\frac{1}{6}&\frac{1}{2}
\end{array}\right].
\end{equation}

\subsubsection{The case $\alpha=0$}
In the case where $\alpha=0$, we have that
$3\int\limits_{\BH_{i}}^{\BH_{j}}\in \Lambda$ for any branchpoint,
and moreover \cite{bren06} that $\mathcal{A}_{\BH_{1}}\left(
\sum_{k=1}\sp{3}\hat{\infty}_k\right)=0$. Thus in this case we
have
\begin{equation*}\widetilde{\boldsymbol{K}}=
\boldsymbol{\hat{K}}_{\BH_{1}}+\mathcal{A}_{\BH_{1}}\left(
\sum_{k=1}\sp{3}\hat{\infty}_k\right)=
\boldsymbol{\hat{K}}_{\BH_{1}}=
\pi\sp\ast({\boldsymbol{K}}_{\infty_+})-\hat e =
\pi\sp\ast({\boldsymbol{K}}_{\infty_+})
\end{equation*}
which yields
\begin{equation*}\boldsymbol{\hat{K}}_{\BH_{1}}=
\pi\sp\ast({\boldsymbol{K}}_{\infty_+})=
\pi\sp\ast(\left[ \begin{array}{cc}
\frac{1}{2}&\frac{1}{2}\\
\frac{1}{6}&\frac{1}{2}
\end{array}\right])=
\dfrac12\left[\begin{matrix}1&1&1&1\\1&1&1&1\end{matrix}\right].
\end{equation*}
This coincides with the result of \cite{bren06} derived by other
methods.

\section{The Ercolani-Sinha conditions}
Here we shall express the transcendental Ercolani-Sinha
constraints on the curve $\hat{\mathcal{C}}$ as conditions on the
curve ${\mathcal{C}}$ and then describe our strategy to solve
them.

With the ordering of the differentials (\ref{diffxhat}) the
Ercolani-Sinha conditions (\ref{eshit2}) take the form
$$(\boldsymbol{n},\boldsymbol{m})\begin{pmatrix}\hat{\mathcal{A}}\\
\hat{\mathcal{B}}
\end{pmatrix}=-2(0,0,0,1),$$
where
$\widehat{\mathfrak{es}}=\boldsymbol{n}\cdot{\hat{\mathfrak{a}}}+
\boldsymbol{m}\cdot{\hat{\mathfrak{b}}}$. Now substituting
$(\boldsymbol{n},\boldsymbol{m})=(n_0,n_1,n_2,n_3,m_0,m_1,m_2,m_3)$
directly into (\ref{pihatsigma}) and making use of (\ref{abcd},
\ref{imcd}) we may deduce that the Ercolani-Sinha vector takes the
form
$$(\boldsymbol{n},\boldsymbol{m})=(n_0,n,n,n,m_0,m,m,m),$$
and thus $\widehat{\mathfrak{es}}$ is fixed under the spatial
symmetry:
$\sigma(\widehat{\mathfrak{es}})=\widehat{\mathfrak{es}}$. (This
result was obtained more generally via a different argument in
\cite{bra10}.) With this simplification we find the remaining
equations encoded in the Ercolani-Sinha conditions take the form
$$(n_0,3n)\begin{pmatrix}x_0& y_0 \\ x_1&y_1\end{pmatrix}+
(m_0,3m)\begin{pmatrix}X_0&Y_0 \\ X_1&Y_1 \end{pmatrix}=
-2\,(0,1).$$ Now using (\ref{abcd}) we have that
\begin{align*}
\begin{pmatrix}X_0& Y_0 \\ X_1&Y_1 \end{pmatrix}
\begin{pmatrix}x_0& y_0 \\ x_1&y_1\end{pmatrix}^{-1}&=
\begin{pmatrix}3& 0 \\ 0&1\end{pmatrix}
\begin{pmatrix}a/3& b \\ b&c+2d \end{pmatrix}
=\begin{pmatrix}3& 0 \\ 0&1\end{pmatrix}\, \tau.
\end{align*}
Upon noting (\ref{diffquots}) and that the periods of
$\hat{\mathcal{A}}$ and $\hat{\mathcal{B}}$ were constructed from
${\hat{\mathbf{u}}_{\ast}}$ we obtain
\begin{theorem} \label{ESHYP} The Ercolani-Sinha constraint on the curve
$\hat{\mathcal{C}}$ yields the constraint
\begin{equation}\label{esgenus2}
(n_0,3n,3 m_0,3m
)\begin{pmatrix}\mathcal{A}\\
\mathcal{B}\end{pmatrix}=-2(0,1)
\end{equation}
on the curve $\mathcal{C}$ with respect to the differentials
$-dx/{(3y)}$, $-xdx/{(3y)}$ and the homology basis
$\{{\mathfrak{a}}_0,{\mathfrak{b}}_0 ,{\mathfrak{a}}_1 ,
{\mathfrak{b}}_1,\}$.
\end{theorem}
We remark also that we have
$$\boldsymbol{\widehat U}=\pi\sp\ast(\boldsymbol{U}),\qquad
\boldsymbol{U}=\frac12(\frac{n_0}3,n)+\frac12(m_0,m)\tau.$$
If we define the cycle
\begin{equation}\label{cyccdef}\boldsymbol{\mathfrak{c}}:=\pi(\widehat{\mathfrak{es}})
=n_0\mathfrak{a}_0+3n{\mathfrak{a}}_1
+3m_0{\mathfrak{b}}_0+3m{\mathfrak{b}}_1
\end{equation} then the
Ercolani-Sinha constraints may be alternately expressed as
\begin{equation}\label{altes}
6\beta_0=\oint\limits_{\widehat{\mathfrak{es}}}
\pi\sp\ast\left(\beta_0 {\mathbf{u}}_{2}+\beta_1
{\mathbf{u}}_{1}\right)= \oint\limits_{\boldsymbol{\mathfrak{c}}}
\left(\beta_0 {\mathbf{u}}_{2}+\beta_1 {\mathbf{u}}_{1}\right).
\end{equation}

At this stage then we have reduced the Ercolani-Sinha constraints
on the curve $\hat{\mathcal{C}}$ to analogous conditions on the
curve ${\mathcal{C}}$. We now use the approach outlined in the
introduction. We know from \cite{bren06} the values of
$(\boldsymbol{n},\boldsymbol{m})$ of the Ercolani-Sinha vector for
the curve (\ref{tmcurve}) for both signs. After changing from the
homology basis of that work to that of the present paper we obtain
for the two cases of (\ref{tmcurve}),
\begin{equation}\label{nvalsc}
(n_0,n,m_0,m)=\begin{cases} (4,1,-3,1)&\qquad+5\sqrt{2},\\
(5,1,-3,0)&\qquad -5\sqrt{2},\end{cases}
\end{equation}
and so we know the cycle $\boldsymbol{\mathfrak{c}}$ for each of
the two loci associated to the tetrahedrally symmetric monopoles.
We also remark that just as
$\tau_\ast(\widehat{\mathfrak{es}})=-\widehat{\mathfrak{es}}$
\cite{hmr99} we also have that
$\tau'_\ast(\boldsymbol{\mathfrak{c}})=-\boldsymbol{\mathfrak{c}}$.
Thus for the $+5\sqrt{2}$ values above and using (\ref{MtauC})
appropriate to this we have $(4,3,-9,3)M_{\tau'}=-(4,3,-9,3)$ and
similarly for $-5\sqrt{2}$ we have
$(5,3,-9,0)M_{\tau'}=-(5,3,-9,0)$.

Now if we now make a change of variable
$$x=\beta\sp{1/3}\,X,\qquad y=\beta\, Y,\qquad
a=\frac{\alpha}{\beta\sp{2/3}},\qquad g=\frac{\gamma}{\beta}$$ then
\begin{align*}
Y^2=(X^3+a\, X+g)^2+4,\qquad
{\mathbf{u}}_{1}=\frac{\de x}{y}=\beta\sp{-2/3}\frac{\de X}{Y}, \quad
{\mathbf{u}}_{2}=\frac{x\;\de x}{y}=\beta\sp{-1/3}\frac{X\de X}{Y},
\end{align*}
and the Ercolani-Sinha constraints take the form
\begin{align}
0&=\oint\limits_{\boldsymbol{\mathfrak{c}}}\frac{\de X}{Y},\label{esreduced1}\\
6\beta\sp{1/3}&=\oint\limits_{\boldsymbol{\mathfrak{c}}}\frac{X\de X}{Y}.
\label{esreduced2}
\end{align}
(We denote by $\boldsymbol{\mathfrak{c}}$ the cycle for both the
scaled and unscaled curves.) The first of these equations may be
viewed as defining $g=g(a)$, and then for this solution the second
gives us $\beta=\beta(g)$. Thus solving the Ercolani-Sinha
constraints has reduced to determining the relation between $g$
and $a$ given by (\ref{esreduced1}). Thus to solve the
Ercolani-Sinha constraints we need to be able to compute periods
of these hyperelliptic integrals. We shall do this numerically
using a variant of the arithmetic-geometric mean used to rapidly
and accurately compute periods of elliptic integrals. We shall
turn to this in the next section.

\section{The AGM method}
In this section we shall recall the connection of the
arithmetic-geometric mean (AGM) to evaluating elliptic integrals
and Richelot's generalisation of this to the genus two setting.
This latter work has been most studied in the setting where the
hyperellptic curve has real branch points and we shall need to
extend this discussion to the case with pairs of complex conjugate
branch points relevant to the monopole setting.
\subsection{AGM: the elliptic case}\label{sagmell}
While the origin of the AGM method dates back to Lagrange it was
Gauss who truly initiated its investigation. A large part of what
is known today seems to be due (or at least known) to him (for
historical notes see \textit{e.g.} \cite{cox}). Let $a\ge b$ be
positive real numbers. The \enf{arithmetic-geometric mean} of
these numbers, denoted $M(a,b)$, is the common limit of the
sequences defined as follows:
\begin{align}
a_{0}&=a, & b_{0}&=b, \nonumber\\
a_{n+1}&=\frac{a_{n}+b_{n}}{2}, & b_{n+1}&=\sqrt{a_{n}b_{n}}.
\end{align}
These two sequences satisfy
\begin{equation*}
a_0\geq a_1\geq\ldots\geq a_n\geq a_{n+1}
\geq\ldots\geq b_{n+1}\geq b_n\geq\ldots\geq b_1\geq b_0 ,
\end{equation*}
which ensures the existence of a common limit
\begin{equation*}
\lim_{n\to\infty}a_{n}=\lim_{n\to\infty}b_{n}=M(a,b).
\end{equation*} Indeed
\begin{equation*}
 a_{n+1}-b_{n+1}\leq a_{n+1}-b_n=\frac12(a_n-b_n)
\end{equation*}
whence \[ 0\leq a_n-b_n \leq 2^{-n}(a-b),  \] which ensures rapid
convergence (which is relevant in the present work).

The remarkable observation of Gauss was the connection of the
Arithmetic-Geometric Mean with elliptic integrals.
\begin{theorem}\enf{AGM \cite{gauss99}}\label{GAGM}
Let $a,b\in\mathbb{R}_{+}$ and let $M(a,b)$ be their arithmetic
geometric mean, then
\begin{equation*}
\int_{0}^{\pi/2}\frac{\mathrm{d}\phi}{\sqrt{a^2\cos^2\phi+b^2\sin^2\phi}
}=\frac{\pi}{2M(a,b)}
\end{equation*}
\end{theorem}
The theorem may be understood in terms of maps
$G_n:\mathcal{E}_{n}\rightarrow\mathcal{E}_{n+1}$ between the
elliptic curves
\begin{equation}\label{agmell}
\mathcal{E}_{n}:\qquad y_n^2=x_n(x_n-a_n^2)(x_n-b_n^2).
\end{equation} Using Theorem \ref{GAGM} all
elliptic integrals of the form $\int_{a}^{b}\frac{\de
x}{\sqrt{P(x)}}$ can be expressed in terms of an appropriate
arithmetic geometric mean after various change of variables. We
also remark that the restriction $a,b\in\mathbb{R}_{+}$ may be
extended to $a,b\in\mathbb{C}\sp\ast$, $a\ne\pm b$, with further
discussion of the square roots taken in the geometric mean. We
shall not need this extension here.
\subsection{Richelot and Humbert: the genus 2 case.}
Richelot \cite{richelot1,richelot2} extended to the hyperelliptic
case  Gauss' connection of the AGM with elliptic integrals.
Humbert \cite{humbert} later gave another view of this
re-interpreting Richelot's findings in terms of the duplication
formulae of 2-variable theta functions, \textit{i.e.~} isogenies
(of type (2,2) ) on Abelian surfaces. We will follow here the
modern exposition of Richelot's work given by Bost and Mestre in
\cite{bm} which describes Richelot's ``changes of coordinates'' in
terms of a correspondence (see below).

At the outset we note that the Richelot-Humbert construction is
only given for the case where the genus 2 curve (represented as a
two-sheeted cover of $\mathbb{P}^{1}$ with six branchpoints) has
all \enf{real} branchpoints. This manifests itself in what follows
by using the ordering of the reals. Just as the AGM of two complex
numbers is correspondingly more complicated than the real setting
the implementations of the Richelot-Humbert construction do not
apply in a straightforward fashion to the case of a genus 2 curve
with complex roots. In the next section we describe the
generalisation needed to apply this for our monopole curve.

Consider the genus 2 curve $\mathcal{C}$
\begin{align}\begin{split}\label{eqcondc}
&\hskip4.5cm  y^2+P(x)Q(x)R(x)=0,\\
&P(x)=(x-a)(x-a'),\quad
Q(x)=(x-b)(x-b'),\quad
R(x)=(x-c)(x-c'),
\end{split}
\end{align}
where the real roots $a,a',b,b',c,c'$ are  ordered as
\[ a<a'<b<b'<c<c' .\]
We may associate to this triple of (real) polynomials $(P,Q,R)$
another triple, $(U,V,W)$, defined by
\begin{align}\label{eqcondc1}
U(x)&=[Q(x),R(x)], &
V(x)&=[R(x),P(x)], &
W(x)&=[P(x),Q(x)].
\end{align}
where $[f,g]:=\cfrac{\mathrm{d}f(x) }{\mathrm{d}
x}g(x)-\cfrac{\mathrm{d}g(x) }{\mathrm{d} x}f(x)$. The roots of
the (quadratic) polynomials $U,V,W$ are all real. If we set
$u<u'$, $v<v'$, $w<w'$ the roots of $U,V,W$ respectively, then one
finds
\begin{equation}\label{relroots}
a \leq v \leq w \leq a'\leq  b \leq w' \leq  u \leq  b' \leq  c \leq  u' \leq  v' \leq  c' .
\end{equation}
(Explicit expressions for the roots of $U,V,W$ will be given below
from which these inequalities can be proven.) Humbert gave a
geometric perspective on this construction. Let $p_{i},p_{i}'$ be
the roots of (the quadratic) polynomial $P_{i}$ ($i=1,2,3$). We
may view the six branch points $\{p_i,p_{i}'\}\in\mathbb{P}^{1}$
as six points on a conic $\mathcal{Q}$. Now given a conic and six
points lying on this we may construct six further points as
follows. Consider the lines $L_i:=\overline{p_{i}p_{i}'}$. The
three lines $L_i$ form a triangle and the new points are the
points of tangency to $\mathcal{Q}$ from the vertices of this
triangle. These are the roots of $[P_{1},P_{2}]$. This is
illustrated in Figure \ref{roots} below.
\begin{figure}[b!]
\begin{center}
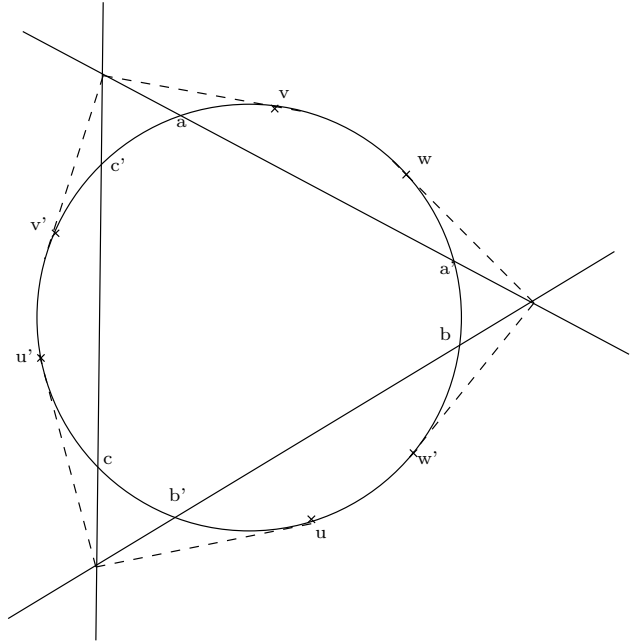
\caption{Roots of $P,Q,R$ and $U,V,W$}\label{roots}
\end{center}
\end{figure}

We thus have a situation similar to Gauss' AGM case: to each pair
of branchpoints one can associate another pair of points which are
closer than the initial ones, and we expect a relation between the
integrals of corresponding pairs. Iterating this process one shows
for every pair the existence of a limit and obtains an expression
for the integrals in terms of these limits. The relation between
integrals suggested and proven by Humbert is
\begin{equation}\label{corr}
\int_{a}^{a'}\frac{S(x)}{\sqrt{-P(x)Q(x)R(x)}}\de x =
2\sqrt{\Delta} \int_{v}^{w}\frac{S(x)}{\sqrt{-U(x)V(x)W(x)}}\de x,
\end{equation}
and similarly for the integrals between the other pairs of
branchpoints. Here $\Delta$ is the determinant of the matrix whose
entries are the coefficients of $P,Q,R$ in the basis
$(1,x,x^{2})$, and $S(x)$ is a polynomial of degree at most one.

There is, however, an important element of difference between the
elliptic and the hyperelliptic cases. The map between elliptic
curves (\ref{agmell}) whose iteration leads to Theorem \ref{GAGM}
is replaced by a \enf{correspondence} in the hyperelliptic
setting. A correspondence $T:\mathcal{C}\to \mathcal{C}'$ of
degree $d$ between two curves $\mathcal{C}$ and $\mathcal{C}'$
associates to every point $p\in \mathcal{C}$ a divisor $T(p)$ of
degree $d$ in $\mathcal{C}'$, varying holomorphically with $p$
\cite{gh}. A correspondence can be presented by its ``curve of
correspondence'', $\mathcal{Z}=\{ (p,q): q\in T(p)\} \subset
\mathcal{C}\times\mathcal{C}'$. In the case of the Humbert
construction, the two curves are
$$\mathcal{C}:\ y^{2}+P(x)Q(x)R(x)=0,\qquad
\mathcal{C}':\ \Delta y'^{2}+U(x')V(x')W(x')=0.$$ The
correspondence between $\mathcal{C}$ and $\mathcal{C}'$ considered
by Humbert  \cite{humbert} is of degree 2, and is given by the
curve $\mathcal{Z}\subset\mathcal{C}\times\mathcal{C}'$ of
equations
\begin{equation}\label{corrcurve}
\begin{cases}
P(x)U(x')+Q(x)V(x')=0, \\
yy'=P(x)U(x')(x-x').
\end{cases}
\end{equation}

In analogy with the pull-back of a map one can also introduce for
correspondences a linear map $\delta_{\mathcal{Z}}:
\Omega^{1}(\mathcal{C}')\to \Omega^{1}(\mathcal{C})$. Then eq.
\eqref{corr} can be interpreted as the relation between
differentials
\begin{equation}\label{corr1}
\delta_{\mathcal{Z}}\left(  \frac{S(x')}{y'} \de x' \right)= \frac{S(x)}{y} \de x,
\end{equation}
together with an analysis of the image of the path joining $a,a'$
(resp. $b,b'$ or $c,c'$) under the correspondence. In fact this
analysis, while mentioned in passing in \cite{bm}, is rather a
crucial point in the extension of Richelot result to the case of
complex conjugate roots. Even if one simply integrates on the
right hand side along a straight line connecting two branchpoints,
the image of this contour via the correspondence \eqref{corrcurve}
may be considerably more complicated. Indeed one can obtain
nontrivial homology cycles in the image and eq. \eqref{corr}
should be interpreted as an identity in $\Jac(\mathcal{C}')$, the
Jacobian of $\mathcal{C}'$. One finds that certain half-periods in
$\mathcal{C}$ are sent to periods in $\mathcal{C}'$ via the
correspondence.

With this background we may now state a version of the
Arithmetic-Geometric Mean for genus 2 curves (with real branch
points).

\subsubsection{The AGM method for genus 2 curves}\label{agmg2th}
Consider the genus 2 curve \eqref{eqcondc}. Define six sequences
$(a_{n})$, $(a'_{n})$, $(b_{n})$, $(b'_{n})$, $(c_{n})$,
$(c'_{n})$ recursively by the conditions:
\begin{itemize}
\item $a_{0}=a$,\; $a'_{0}=a'$,\; $b_{0}=b$,\; $b'_{0}=b'$,\;
    $c_{0}=c$,\; $c'_{0}=c'$;
\item $a_{n+1},a'_{n+1},b_{n+1},b'_{n+1},c_{n+1},c'_{n+1}$
    \;are roots of $U_{n}V_{n}W_{n}$, ordered as follows
\begin{align}\label{orderrroots}
a_{n+1}<a'_{n+1}<b_{n+1}<b'_{n+1}<c_{n+1}<c'_{n+1},
\end{align}
\end{itemize}
where, for every $n$,
\begin{align*}
P_n(x)=(x-a_n)(x-a'_n), \;\;
Q_n(x)=(x-b_n)(x-b'_n),\;\;
R_n(x)=(x-c_n)(x-c'_n),\\
U_n(x)=[Q_n(x),R_n(x)],\;\;
V_n(x)=[R_n(x),P_n(x)], \;\;
W_n(x)=[P_n(x),Q_n(x)].\qquad
\end{align*}

Bost and Mestre \cite{bm} give an explicit expression for these
sequences:
\begin{align}\begin{split}\label{sequences}
a_{n+1}&=\frac{c_{n}c_{n}'-a_{n}a_{n}'-B_n  }{c_n+c_n'-a_n-a_n'},\quad
a_{n+1}'=\frac{b_{n}b_{n}'-a_{n}a_{n}'-C_n  }{b_n+b_n'-a_n-a_n'},\\
b_{n+1}&=\frac{b_{n}b_{n}'-a_{n}a_{n}'+C_n  }{b_n+b_n'-a_n-a_n'},\quad
b_{n+1}'=\frac{c_{n}c_{n}'-b_{n}b_{n}'-A_n  }{c_n+c_n'-b_n-b_n'},\\
c_{n+1}&=\frac{c_{n}c_{n}'-b_{n}b_{n}'+A_n  }{c_n+c_n'-b_n-b_n'},\quad
c_{n+1}'=\frac{c_{n}c_{n}'-a_{n}a_{n}'+B_n  }{c_n+c_n'-a_n-a_n'},
\end{split}
\end{align}
with
\begin{align*}
A_n&=\sqrt{ (b_n-c_n)(b_n-c_n')(b_n'-c_n)(b_n'-c_n')  },\\
B_n&=\sqrt{ (c_n-a_n)(c_n-a_n')(c_n'-a_n)(c_n'-a_n')  },\\
C_n&=\sqrt{ (a_n-b_n)(a_n-b_n')(a_n'-b_n)(a_n'-b_n')  }.
\end{align*}
These can be derived finding the roots for $U_{n},V_{n}, W_{n}$
as follows
\begin{align}
\begin{split}\label{expruvw}
u_{n},u'_{n}&=\frac{c_{n}c_{n}'-b_{n}b_{n}'\mp A_n  }{c_n+c_n'-b_n-b_n'},\\
v_{n},v_{n}'&=\frac{c_{n}c_{n}'-a_{n}a_{n}' \mp B_n  }{c_n+c_n'-a_n-a_n'},\\
w_{n},w_{n}'&=\frac{b_{n}b_{n}'-a_{n}a_{n}' \mp C_n  }{b_n+b_n'-a_n-a_n'},
\end{split}
\end{align}
and ordering them according to \eqref{orderrroots}. One sees
directly from the expressions above that  $v_{n} \leq w_{n}  \leq
w'_{n} \leq u_{n} \leq  u_{n}' \leq  v_{n}'$ (cf. also eq.
\eqref{relroots}). Thus we set
\begin{equation}\label{rootsintermsofuvw}
a_{n+1}=v_{n}, \quad a_{n+1}'= w_{n},  \quad b_{n+1}=w'_{n},
\quad b_{n+1}'= u_{n}, \quad c_{n+1} =u_{n}', \quad   c'_{n+1}=v_{n}',
\end{equation}
and \eqref{sequences} then follow. We then obtain:

\begin{theorem}[\enf{Richelot \cite{richelot2}, Bost and Mestre \cite{bm}}]\label{richthm}
With the  above definitions, the sequences  $(a_{n})$, $(a'_{n})$,
$(b_{n})$, $(b'_{n})$, $(c_{n})$, $(c'_{n})$ converge pairwise to
common limits
\begin{align*}
&\lim_{n\to\infty} a_n=\lim_{n\to\infty} a_n'=\alpha\equiv M(a,a'),\\
&\lim_{n\to\infty} b_n=\lim_{n\to\infty} b_n'=\beta\equiv M(b,b'),\\
&\lim_{n\to\infty} c_n=\lim_{n\to\infty} c_n'=\gamma\equiv M(c,c').
\end{align*}
Furthermore, for any polynomial $S(x)$ of degree at most one, the
following relations hold:
\begin{align}\begin{split}\label{intthm}
I(a, a')\equiv\int_{a}^{a'}\frac{S(x)\mathrm{d}x}{\sqrt{-P(x)Q(x)R(x)  }}
=\pi T \frac{S(\alpha)}{(\alpha-\beta)(\alpha-\gamma)},\\
I(b, b')\equiv\int_{b}^{b'}\frac{S(x)\mathrm{d}x}{\sqrt{-P(x)Q(x)R(x)  }}
=\pi T \frac{S(\beta)}{(\beta-\alpha)(\beta-\gamma)},\\
I(c, c')\equiv\int_{c}^{c'}\frac{S(x)\mathrm{d}x}{\sqrt{-P(x)Q(x)R(x)  }}
=\pi T \frac{S(\gamma)}{(\gamma-\alpha)(\gamma-\beta)},
\end{split}
\end{align}
where
\begin{equation}\label{tnt}
T=\prod_{n=0}^{\infty}t_n,\quad  t_n=
\dfrac{2\sqrt{\Delta_n}}{\sqrt{(b_n+b_n'-a_n-a_n')(c_n+c_n'-b_n-b_n')(c_n+c_n'-a_n-a_n')}}.
\end{equation}
\end{theorem}\vspace*{4mm}

The proof of the convergence of the sequences $(a_{n})$,
$(a'_{n})$ (and likewise $(b_{n})$, $(b'_{n})$; $(c_{n})$,
$(c'_{n})$) is similar to that of the elliptic case. Using
\eqref{corr} we find
\begin{align}\begin{split}\label{eqintn}
\int_{a_{n}}^{a_{n}'}\frac{S(x)}{ \sqrt{ -P_{n}Q_{n}R_{n} } }\de x =& \;
2\sqrt{\Delta_{n}}\int_{a_{n+1}}^{a_{n+1}'}\frac{S(x)}{ \sqrt{ -[P_n,Q_n][Q_n,R_n][R_n,P_n] } }\de x\\
=& \;t_{n}\int_{a_{n+1}}^{a_{n+1}'}\frac{S(x)}{ \sqrt{ -P_{n+1}Q_{n+1}R_{n+1} } }\de x,
\end{split}\end{align}
and the relations \eqref{intthm} follow upon taking the limit for
$n\to\infty$ and using the residue theorem. Integrals between
other pairs of branchpoints (\textit{e.g.} $a'$ and $b$) may also
be calculated using the same method in conjunction with
appropriate fractional linear transformations. We remark that the
integral between $b$ and $b'$ given above has opposite sign to
that in \cite{bm}, because of a different choice of
conventions\footnote{Bost and Mestre, in their note 2, p. 51 of
\cite{bm}, claim that they want to recover the ``classical
identity'' $I_{a}-I_{b}+I_{c}=0$. With our choice of convention
for sheets, the relation between integrals become
$I_{a}+I_{b}+I_{c}=0$, which follows from the fact that the
integral around a cycle encircling all the cuts, oriented so that
the upper arc goes from negative to positive real values, is
zero.}.

\subsection{Generalisation to the genus 2 case with complex conjugate roots}
We now generalise Richelot's method of the previous section to the
case where the branchpoints are not all real but the polynomials
$P,Q,R$ are still real. This corresponds to the three pairs of
complex conjugate branchpoints,
\begin{align*}
a'=\bar{a},\quad b'=\bar{b},\quad c'=\bar{c},
\end{align*}
the case relevant for our monopole curve \eqref{quotientcurve}. We
further order the roots such that
$$\Re(a)=\Re(a') < \Re(b)=\Re(b') < \Re(c)=\Re(c')$$
and for definiteness take  $\Im(a)<0$, $\Im(b)<0$, $\Im(c)<0$.
This splitting into complex conjugate pairs was given for the
quotient monopole curve in \eqref{bpquotient} and
\eqref{bpquotientcc}.

Of course all of the polynomial relations given in the previous
sections extend to the case of arbitrary complex branchpoints and
so the relation (\ref{corr1}) between the differentials on
$\mathcal{C}$ and $\mathcal{C}'$ still holds true for this case.
The difference with complex branch points arises at two points.
First, with complex roots, there is no natural way to order the
branchpoints and hence no natural way of splitting the
branchpoints into pairs; thus there is choice in constructing a
sequence of branchpoints to iterate. (This same feature is present
with the ordinary AGM when the elliptic curve does not have real
structure.) Second, as noted earlier, the image of the path
between branchpoints under the correspondence may be quite
complicated. Restricting attention to the case of the three
quadratics $P,Q,R$ having complex conjugate roots simplifies the
problem somewhat. Although the initial branchpoints are complex
and a relation analogous to \eqref{relroots} cannot be written,
nevertheless the roots of $U_{0},V_{0},W_{0}$ are real (as can be
seen by considering their explicit expressions in \eqref{expruvw})
and so can be ordered. In contrast with the purely real case
however, this ordering is not unique: in the real case the
ordering of $u$, $u'$, $v$, $v'$, $w$, $w'$ depended only on the
relative ordering of $a$, $a'$, $b$, $b'$, $c$, $c'$ on the real
line; in the complex conjugate case this now depends on their
imaginary parts. Depending on the ordering of $u$, $u'$, $v$,
$v'$, $w$, $w'$, equation \eqref{eqintn} relating the integrals
between  the three pairs of branchpoints on $\mathcal{C}$ and
$\mathcal{C}'$ needs to be modified appropriately. We shall focus
here on the case relevant for the monopole quotient curve.

\subsubsection{The AGM method for the quotient monopole curve}
Consider the quotient monopole curve (\ref{quotientcurve}).
Ordering the branchpoints as in section \ref{properties}  we have
\begin{align}\label{bpcc}
a=B_{4},\; a'=B_{3}; \quad b=B_{5},\; b'=B_{2}; \quad c=B_{6},\; c'=B_{1}.
\end{align}
Calculating  $u$, $u'$, $v$, $v'$, $w$, $w'$ via \eqref{expruvw},
and examining their relative ordering we find the following two
cases:
\begin{align}
\mathrm{\enf{case 1}:}\; \alpha>0, &\qquad v \leq w \leq w' \leq  u \leq   u' \leq  v' ;
\label{case1}\\
\mathrm{\enf{case 2}:} \;\alpha<0, & \qquad  u\leq v \leq w \leq u' \leq  v'  \leq  w'.
\label{case2}
\end{align}

\enf{Case 1.} This is exactly the same situation as considered in
\cite{bm}, Thus \eqref{rootsintermsofuvw} still holds and so the
sequences $a_{n}$, $a'_{n}$, $b_{n}$, $b'_{n}$, $c_{n}$, $c'_{n}$
are still given by \eqref{sequences}. Therefore the first equality
in  \eqref{eqintn} holds in view of \eqref{case1} and so the
integrals between complex conjugate pairs of branchpoints are
still expressed by \eqref{intthm}. We shall return to a discussion
of integrals between other pairs of branchpoints shortly.

\enf{Case 2.} The different ordering of \eqref{case2} means that
\eqref{rootsintermsofuvw} no longer holds for the first step of
the recurrence. We modify this as follows. In view of
\eqref{case2} for $n=1$ we take
 \begin{equation*}
a_{1}= u, \quad a_{1}'= v, \quad b_{1}=w \quad b'_{1}=u', \quad  c_{1}=v', \quad c_{1}'= w'.
\end{equation*}
Now at this stage the curve $\mathcal{C}'$ has all real
branchpoints and hence the Richelot-Humbert iteration can be
applied as previously. Thus the only change in this case occurs at
the first step of the recurrence. Let us denote by $I(p,q)$ the
integral on $\mathcal{C}$ between $p,q$ on the first sheet, and by
$I^{(i)}$ the integrals on the curve $\mathcal{C}^{(i)}$ of
equation $y^{2}+P_{i}(x)Q_{i}(x)R_{i}(x)=0$ on the same sheet. The
integrals $I^{(i)}$ can be then expressed by equations
\eqref{intthm}, using the AGM method for the curve
$\mathcal{C}\sp{(1)}$. We obtain an expression for the integrals
$I(p,q)$ on $\mathcal{C}$ as follows. Numerically\footnote{More
specifically, we study the images on the curve $\mathcal{C}'$ of
equation $\Delta y^{2}+U(x)V(x)W(x)=0$, in order to understand the
first equality of \eqref{eqintn}  (as the second follows
immediately from the first). We find, for instance, that the
straight line between $a$ and $a'$ on $\mathcal{C}$, call it
$\gamma(a,a')$, is sent to a closed cycle encircling $a'_{1}$ and
$b_{1}$ on $\mathcal{C}'$. Noticing that in the second equality of
\eqref{eqintn} there is a factor of $1/2$, absorbed in the
definition of $t_{i}$, we obtain that the image of $\gamma(a,a')$
on the curve $y^{2}+P_{1}(x)Q_{1}(x)R_{1}(x)=0$, \textit{i.e.~}
$\mathcal{C}'$, is precisely the path from $a'_{1}$ and $b_{1}$.}
we calculate the images on $\mathcal{C}'$ under the correspondence
\eqref{corrcurve} of the straight line contours of integration
between the branchpoints of $\mathcal{C}$. The resulting contours
of integration on the right hand side of \eqref{intcorr2} yield
\begin{align}\begin{split}\label{intcorr2}
I(a,a')=& \; t_{0}I^{'}(a_{1}',b_{1}),\\
I(b,b')=&\; t_{0}I^{'}(c'_{1},a_{1})= \;t_{0}(-I^{'}(a_{1}',b_{1}) +I^{'}(b_{1}',c_{1})),\\
I(c,c')=&-t_{0} I^{'}(b_{1}',c_{1}).
\end{split}
\end{align}
A similar numerical analysis of the images of the paths between
other pairs of branchpoints then yields
\begin{align}
\begin{split}\label{otherint}
I(a,b)=& \,\frac{1}{2}t_{0}\,(I^{'}(a_{1},a'_{1}) + I'(b_{1}',c_{1})),\\
I(a',b')=&\,\frac{1}{2}t_{0}\,(I^{'}(a_{1},a'_{1}) - I'(b_{1}',c_{1})),\\
I(b,c)=& \,\frac{1}{2}t_{0}\,(-I^{'}(a_{1},a'_{1}) +I'(a_{1}',b_{1})-I'(b_{1},b'_{1})),\\
I(b',c')=& \,\frac{1}{2}t_{0}\,(-I^{'}(a_{1},a'_{1}) -I'(a_{1}',b_{1})-I'(b_{1},b'_{1})).
\end{split}\end{align}
Recalling that we are able to express the integrals $I'(p,q)$
applying the AGM method to the curve $\mathcal{C}'$ (with real
branchpoints) using Theorem \ref{richthm} and earlier remarks we
are able then to calculate all integrals between branchpoints on
$\mathcal{C}$.

Finally, we can apply similar numerical considerations to the
integrals between non complex conjugate branchpoints  in \enf{case
1} to obtain
\begin{align}
\begin{split}\label{intalphaneg}
I(a,a')=& \; t_{0}I^{'}(a_{1},a_{1}'),\\
I(b,b')=&\; t_{0}I^{'}(b_{1},b'_{1}),\\
I(c,c')=&\;t_{0} I^{'}(c_{1},c'_{1}),\\
I(a,b)=& \frac{1}{2}\,t_{0}\,(\; I'(a_{1}',b_{1})-I'(c_{1},c'_{1})),\\
I(a',b')=&\frac{1}{2} \,t_{0}\,( I'(a_{1}',b_{1})+I'(c_{1},c'_{1})),\\
I(b,c)=& \frac{1}{2}\,t_{0}\,(\;-I^{'}(b_{1},b'_{1}) -I'(b_{1}',c_{1})-I'(c_{1},c'_{1})),\\
I(b',c')=&\frac{1}{2} \,t_{0}\,(I^{'}(b_{1},b'_{1}) +I'(b_{1}',c_{1})+I'(c_{1},c'_{1}))).
\end{split}\end{align}
\section[Solving the Ercolani-Sinha constraints via the
AGM method] {Solving the Ercolani-Sinha constraints via the
AGM}\label{AGMessect} We have shown how the Ercolani-Sinha
constraints are reduced to finding the
$(a,g):=(\alpha/\beta\sp{2/3},\gamma/\beta)$ such
\begin{equation}\label{agmes}
0=\oint\limits_{\boldsymbol{\mathfrak{c}}}\frac{\de
X}{Y},\qquad Y^2=(X^3+a\, X+g)^2+4
\end{equation} and for the cycle
$\boldsymbol{\mathfrak{c}}$ given by (\ref{cyccdef}) and
(\ref{nvalsc}) for $a>0$ and $a<0$. Our strategy is as follows.
Using the arc expansion \eqref{basisqexp} we may express the cycle
$\boldsymbol{\mathfrak{c}}$ in terms of integrals between
branchpoints. These integrals are then evaluated via the AGM
method of the previous section using (\ref{intthm},
\ref{intalphaneg}) for the case $a>0$ and (\ref{intcorr2},
\ref{intcorr2}) for $a<0$. This has been implemented in Maple. The
advantage of the AGM method is that it is much faster than the
direct numerical integrations between branchpoints as it deals
only with polynomial manipulations; moreover, the convergence of
the sequences $(a_{n})$, $(a'_{n})$, $(b_{n})$, $(b'_{n})$,
$(c_{n})$ only needs very few steps, usually 6 or 7, for the
precision we require. These considerations allow us to
successfully solve the Ercolani-Sinha constraints iteratively, as
described in the next subsection. A useful check of the method is
the calculation of the matrix of $\mathfrak{a}$ and $\mathfrak{b}$
periods. These periods may again be reduced to integrals between
branchpoints using the arc expansion \eqref{basisqexp} and
consequently evaluated via the AGM. We find agreement with the
same quantities being evaluated by other methods, yet with
significant improvement in speed.

\subsection{Numerical solutions}\label{secagmes}

\begin{figure*}
\begin{center}
\includegraphics[height=250pt]{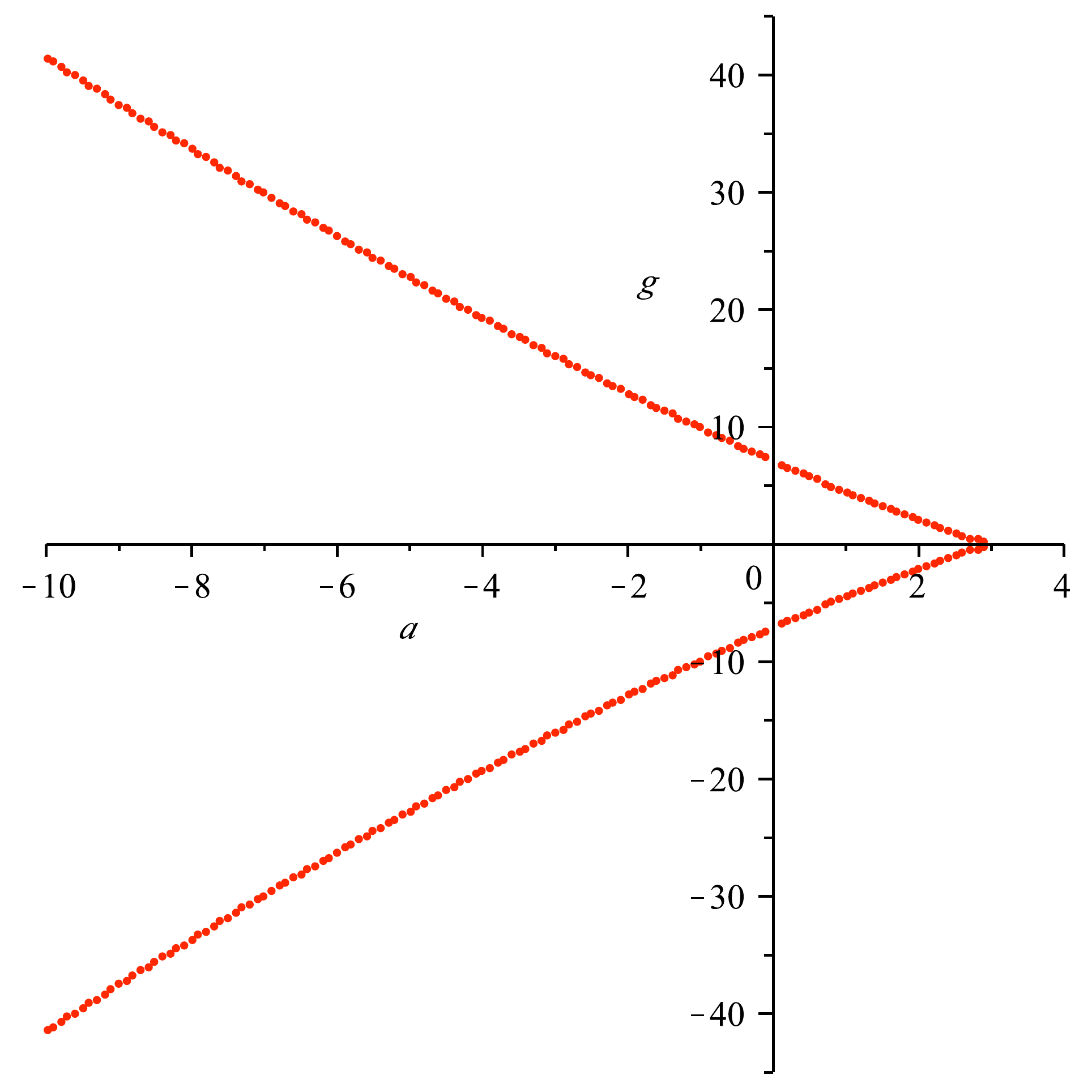}
\caption{Solutions to the Ercolani-Sinha constraints}\label{ESsoln}
\end{center}
\end{figure*}

To begin, we wish to find solutions to \eqref{agmes} starting from
the point  $a=0$ and $g=5\sqrt{2}$. The integral over
$\boldsymbol{\mathfrak{c}}$ is reduced to integrals between
branchpoints as described above. We then proceed iteratively as
follows. We start varying $\alpha$ by a small $\epsilon$, namely
$\alpha_{i}=i\cdot\epsilon$; we then vary $\gamma$ in smaller
steps, $\gamma_{i,k}=5\sqrt{2}+k\cdot\epsilon^{2}$. For every such
pair $(\alpha_{i}, \gamma_{i,k})$ we calculate the periods using
the AGM method, and hence compute the first constraint in
\eqref{agmes}. For every $\alpha_{i}$ we take the $\gamma_{i,k}$
for which this constraint vanishes. We repeat this a sufficiently
large number of times,  obtaining the curve in Figure
\ref{ESsoln}. We have used a step size of $10^{-1}$ for $a\in
(0,2.8)$, and of $10^{-2}$ for $a\in (2.8,3.0)$ to obtain greater
detail in this interval. For values of $a\in(0,3)$, the outcome is
that we have a curve of solutions in the space of parameters
passing through the points $(0,5\sqrt{2})$ up to the point
$(3,0)$. This point does not however belong to the solution curve.
We may also extend this curve for negative values of $a$ to the
left of the point $(0,5\sqrt{2})$. In Figure \ref{ESsoln} we plot
100 points corresponding to $a<0$, $g>0$.

The point $(a,g)=(3,0)$ is in fact a singular point as 4 of the
branchpoints collide pairwise, giving two singular points at $\pm
i$. This results in a rational curve with equation
\begin{equation}\label{ratmon}
y^{2}=(x^{3}+3x)^{2}+4=(x^{2}+4)(x^{2}+1)^{2}.
\end{equation}
Note that the curves of equation \eqref {agmes} with $g=0$ and $a>
0$ are all hyperelliptic with the only exception being precisely
the case $a=3$ above where the curve is reducible. In terms of the
original spectral curve (\ref{curve}) we have $\alpha^3=27\beta^2$
and \cite{hmm95} noted the loci under consideration here being
asymptotic to this at one end. As a rational curve has no
nontrivial cycles, the second of the Ercolani-Sinha constraints
(\ref{esreduced2}) that fixes $\beta$ tells us $\beta=\infty$ and
so our solution curves are asymptotic to the point $(a,g)=(3,0)$.
We remark that the behaviour of the solution curve in Figure
\ref{ESsoln} is consistent with the findings of \cite{sut96b},
where Sutcliffe predicts that the curve \eqref{ratmon}, describing
a configurations of three unit-charge monopoles with dihedral
$\texttt{D}_{3}$ symmetry, constitutes an asymptotic state for a
3-monopole configuration (cf. eq. (4.16) in \cite{sut96b}).

When trying to extend the solution curve to $g<0$, we do not
observe any values of the parameters satisfying \eqref{agmes} with
the first set of integers of \eqref{nvalsc}. But since the point
$(a,g)=(3,0)$ does not belong to the solution curve, continuity
arguments do not prevent us using the second set of integers of
\eqref{nvalsc}. With these we manage to extend the solution curve
through the point $(0,-5\sqrt{2})$, which again corresponds to a
tetrahedral monopole, now with a different orientation. This curve
is also shown in Figure \ref{ESsoln} We point out that the arc of
the curve for $g<0$ is precisely the reflection with respect to
the $a$-axis of the arc obtained for $g>0$. Because of this
symmetry we may focus attention on the case $g>0$ in what follows.

\begin{figure*}
\begin{center}
\includegraphics[height=250pt]{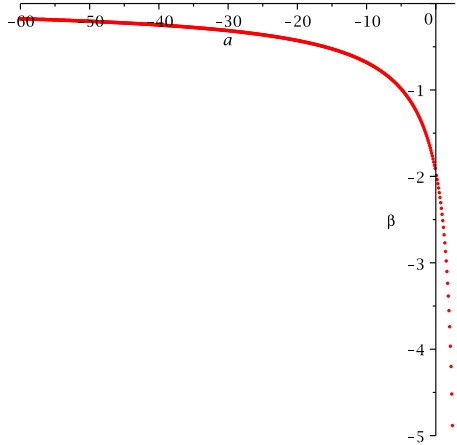}
\caption{$\beta=\beta(a)$}\label{plot_abeta}
\end{center}
\end{figure*}

Having determined the relationship between $a$ and $g$ we then use
this to find $\beta$ via
$$6\beta\sp{1/3}=\oint\limits_{\boldsymbol{\mathfrak{c}}}\frac{X\de
X}{Y}.$$ Again these are just integrals determined via the AGM and
we present $\beta=\beta(a)$ in Figure \ref{plot_abeta}. We shall
interpret these results and compare them with other work in the
next section.

\section{Discussion}
Before discussing our results let us summarise our argument thus
far. In this paper we have constructed the spectral curve
associated to charge three monopoles with cyclic (but not
dihedral) symmetry $\texttt{C}_3$. By imposing cyclic symmetry the
original genus 4 spectral curve $\hat{\mathcal{C}}$ (\ref{curve})
was shown to cover the genus two hyperelliptic curve $\mathcal{C}$
(\ref{quotientcurve}). By making use of a well adapted homology
basis, the theta functions and data appropriate for the monopole
solution were shown to be expressible in terms analogous data for
the quotient curve $\mathcal{C}$. Thus the construction of an
appropriate spectral curve reduced to questions purely in terms of
the curve $\mathcal{C}$. In particular the transcendental
constraints of the Hitchin construction reduce to the single
transcendental constraint
\begin{equation*}
0=\oint\limits_{\boldsymbol{\mathfrak{c}}}\frac{\de
X}{Y}
\end{equation*}
for the scaled curve $Y^2=(X^3+a\, X+g)^2+4$ and a specified cycle
$\boldsymbol{\mathfrak{c}}$. This may be viewed as defining a
function $g=g(a)$ and the monopole curve is determined in terms of
this. The ``special function'' $g(a)$ warrants further study. Here
we have made a numerical study of this. Our numerical study used a
genus 2 extension of the arithmetic-geometric mean found by
Richelot. Like its genus one counterpart Richelot's extension
converges extremely rapidly and is an excellent means of
evaluating such integrals. Richelot's method has (to our
understanding) been used almost entirely in the setting of genus
two curves with real branch points. When extended to the case of
complex (here conjugate) branch points several new features arose.

\begin{figure}
\begin{minipage}[t]{0.48\textwidth}
\centering
\includegraphics[height=200pt]{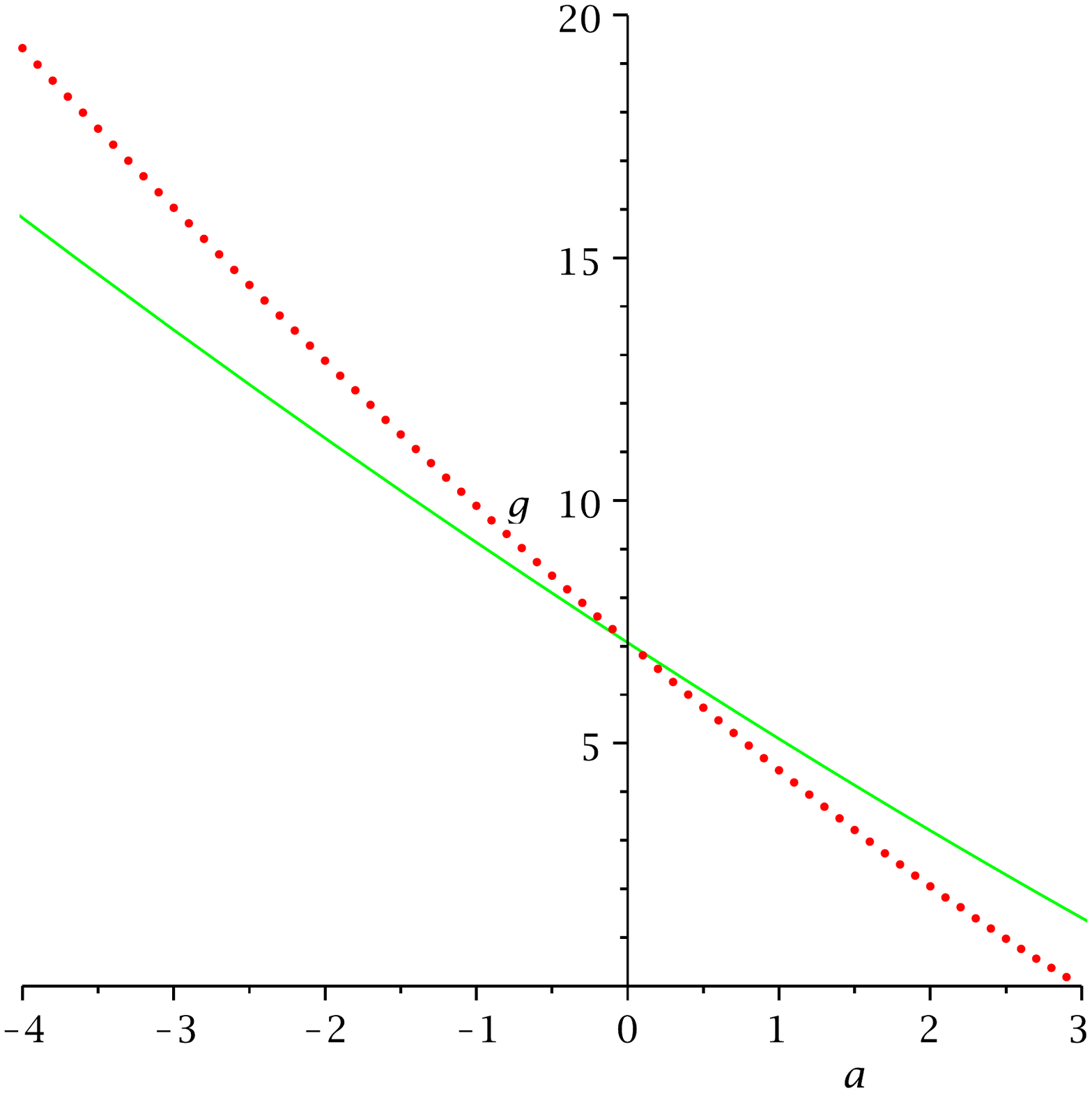}
\caption{$g(a)$ for small values compared with Sutcliffe's approximation (solid line).}\label{avg}
\end{minipage}
\hspace{0.2cm}
\begin{minipage}[t]{0.48\textwidth}
\centering
\includegraphics[height=200pt]{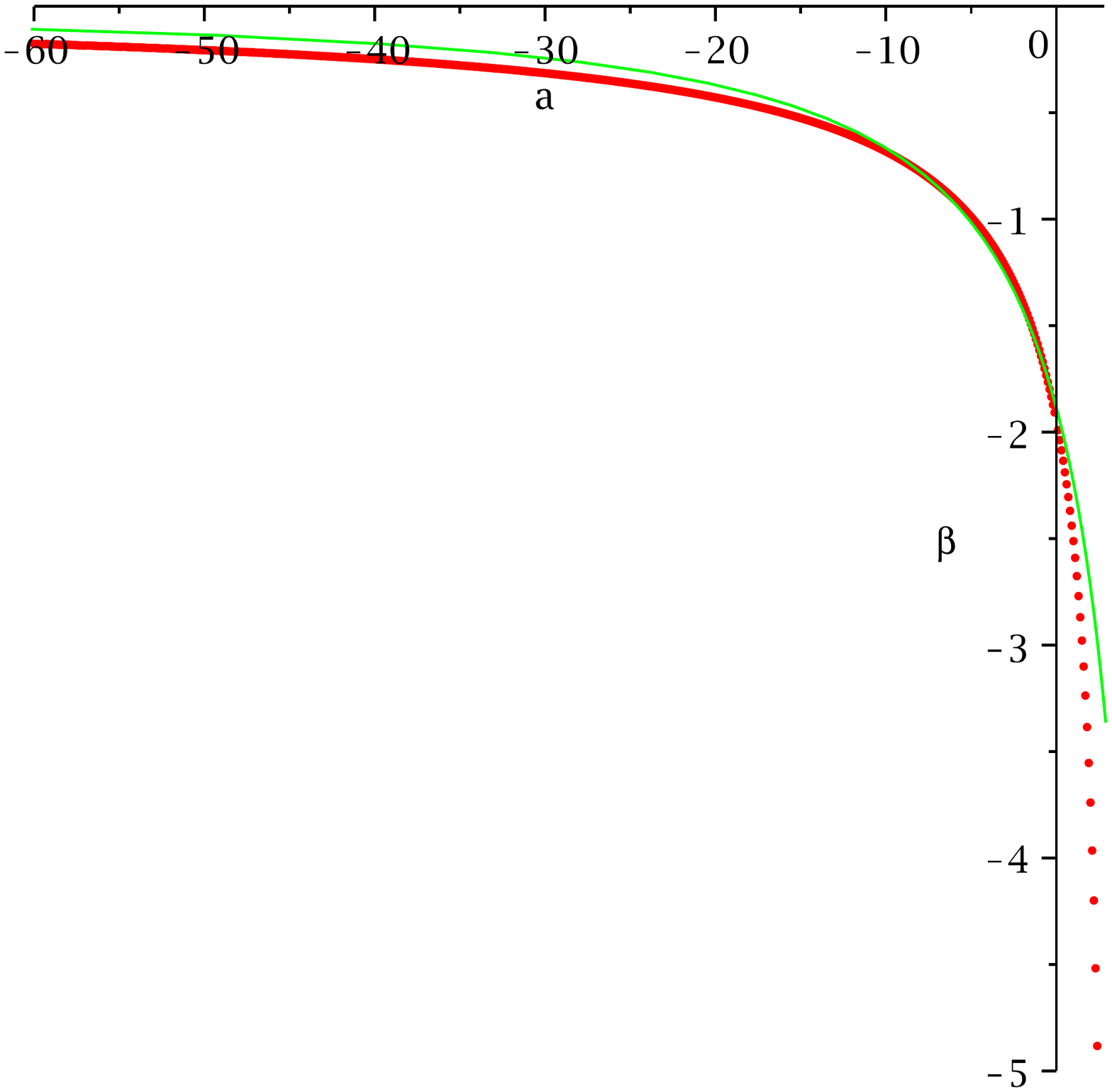}
\caption{$\beta(a)$ compared with Sutcliffe's approximation (solid line).}\label{avb}
\end{minipage}
\end{figure}

\begin{figure}
\begin{center}
\includegraphics[height=200pt]{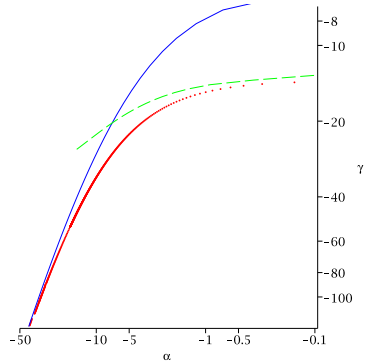}
\caption{A log-log plot of the asymptotic behaviour of $\alpha$ versus
$\gamma$ according to Hitchin, Manton and Murray (solid), Sutcliffe (dash) and
here (dots).}\label{asb2}
\end{center}
\end{figure}

Our results extend work of both Hitchin, Manton and Murray
\cite{hmm95} and Sutcliffe \cite{sut96b} which both describe
cyclically symmetric charge three monopoles. In the former the
following picture of the scattering of three monopoles,
corresponding to geodesic motion along one of our loci is given.
Three unit charge monopoles come in at the vertices of an
equilateral triangle, moving towards its centre, in the
$x_{1}-x_{2}$ plane. Asymptotically this equilateral triangular
configuration corresponds to the reducible spectral curve at
$a=3$. At $a=0$ the three monopoles coalesce instantaneously into
a tetrahedron. Depending on whether the equilateral triangle is
below or above the $x_{1}-x_{2}$ plane we have distinct
orientations of the tetrahedron corresponding to $g>0$ or $g<0$.
Finally the tetrahedron (with say $g>0$) breaks up into a unit
charge monopole moving along the positive $x_{3}$-axis at
$(0,0,b)$ and an axisymmetric charge 2 monopole, moving along the
negative $x_{3}$-axis at $(0,0,-b/2)$. The reducible curve
corresponding to the product of these configurations is
$$
0=(\eta+2b\zeta)(\eta^2-2b\eta\zeta+[b^2+\frac{\pi^2}4]\zeta^2)
=\eta^3+[\frac{\pi^2}4-3b^2]\eta \zeta^2+2b(b^2+\frac{\pi^2}4)\zeta^3
$$
from which we have the asymptotic behaviour at this end of the
scattering given by
\begin{equation}\label{asymp}
\alpha \sim(\pi^2/4 -3b^2),\qquad \gamma \sim
2b(b^2+\pi^2/4),\qquad \beta \sim 0,
\end{equation}
where we have ignored terms vanishing as $b$ tends to infinity.
Sutcliffe investigated the same locus of monopoles numerically
finding approximate twistor data by considering Painlev\'e
analysis of the Nahm data at the pole. This led to approximate
forms of $\alpha$, $\beta$ and $\gamma$ described parametrically.
In Figures \ref{avg}, \ref{avb} we plot this approximate data
alongside the exact results. Despite not actually giving a
spectral curve at any point the energy densities obtained by
Sutcliffe upon solving the Nahm equations qualitatively reflect
the scattering behaviour described above. To compare with the
asymptotic prediction (\ref{asymp}) we must revert to $\alpha$ and
$\gamma$. In Figure \ref{asb2} we give a log-log plot of the exact
values against this asymptotic prediction and alongside that of
Sutcliffe's approximate curve. Our results reproduce the
asymptotic behaviour of (\ref{asymp}). In fact our approach could
be extended to enable the calculation of both analytic and
numerical corrections to this leading behaviour.

\begin{figure}
\begin{minipage}[t]{0.3\textwidth}
\centering
\includegraphics[height=160pt]{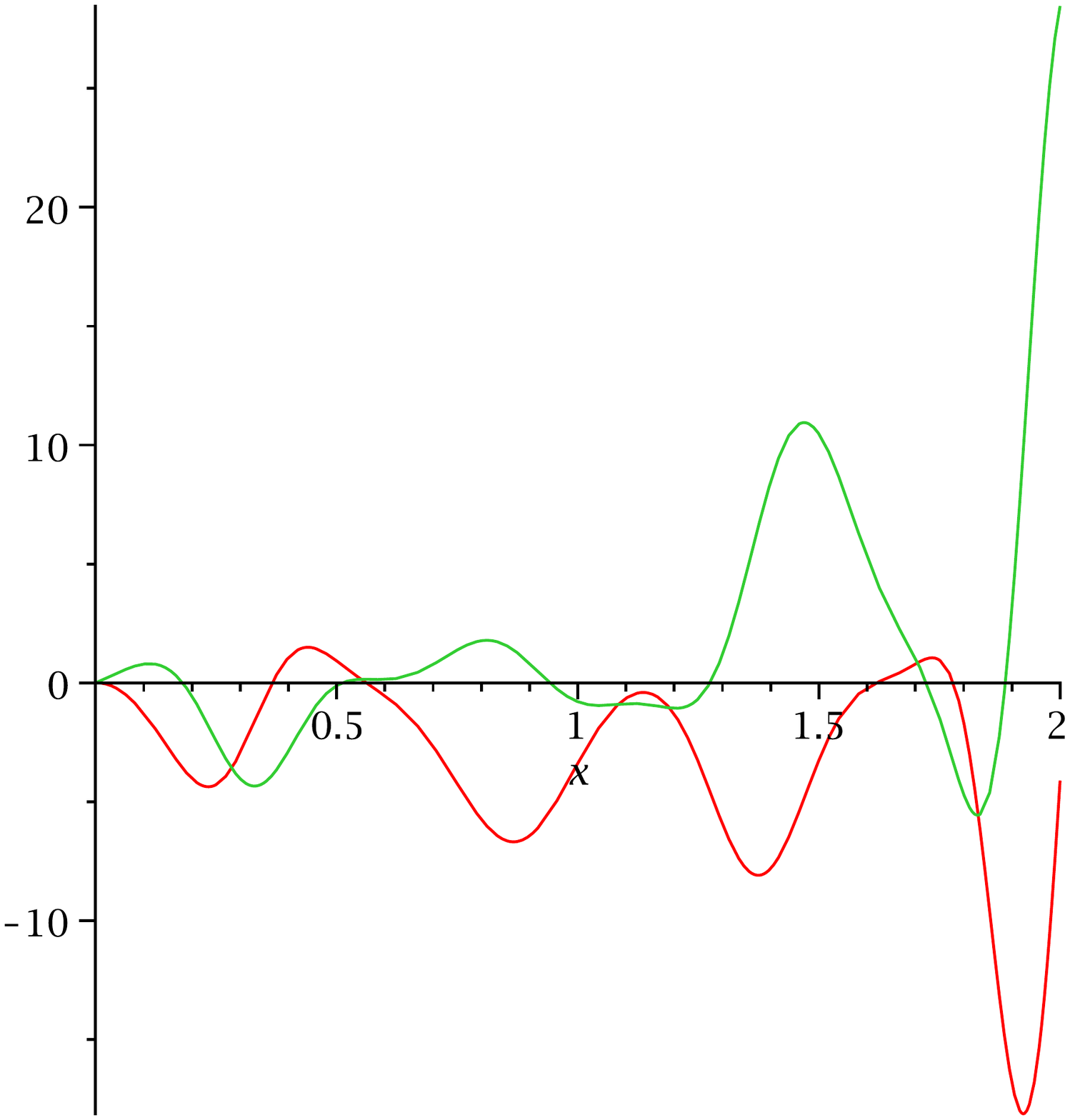}
\end{minipage}
\begin{minipage}[t]{0.3\textwidth}
\centering
\includegraphics[height=160pt]{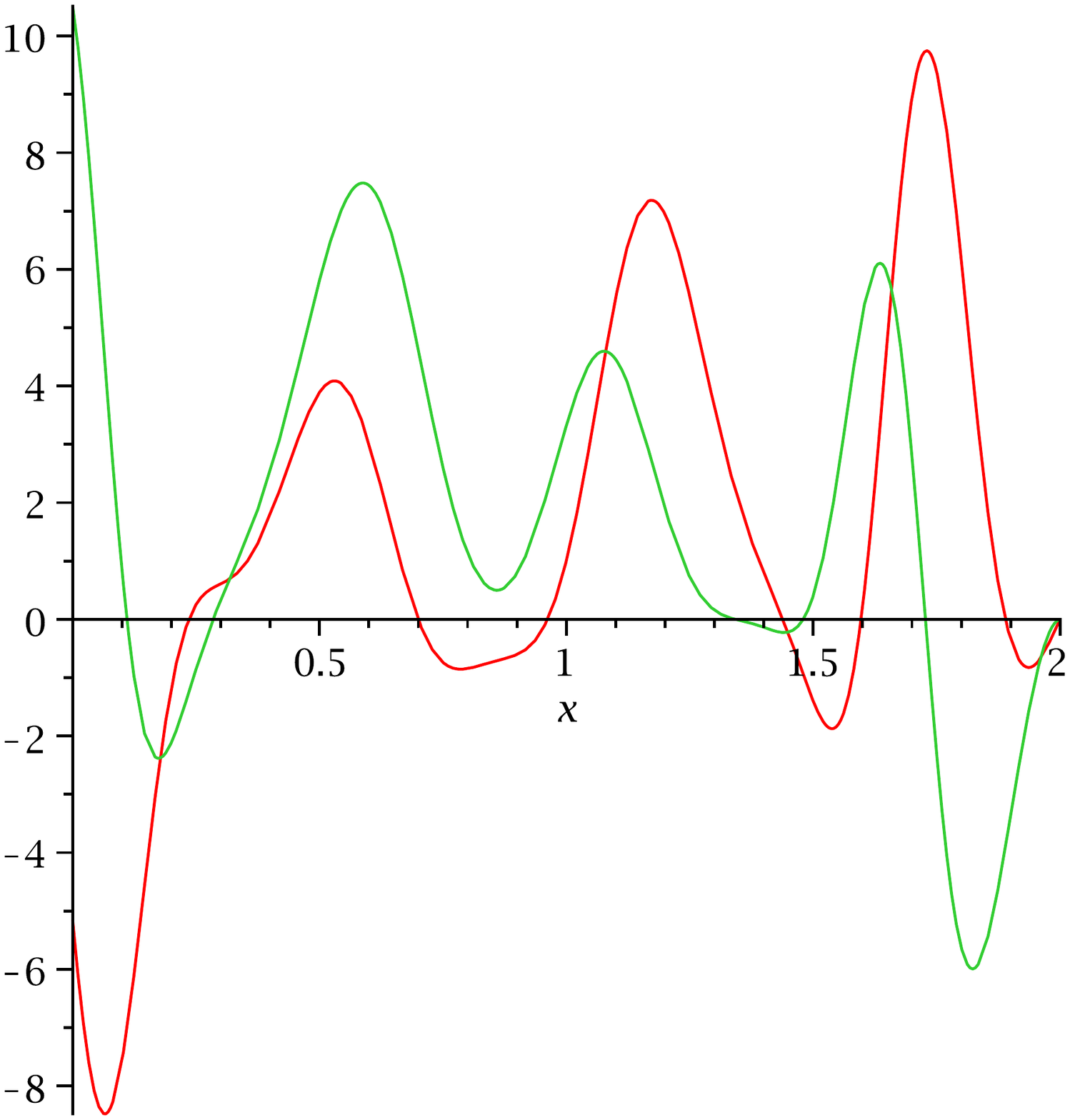}
\end{minipage}
\begin{minipage}[t]{0.3\textwidth}
\centering
\includegraphics[height=160pt]{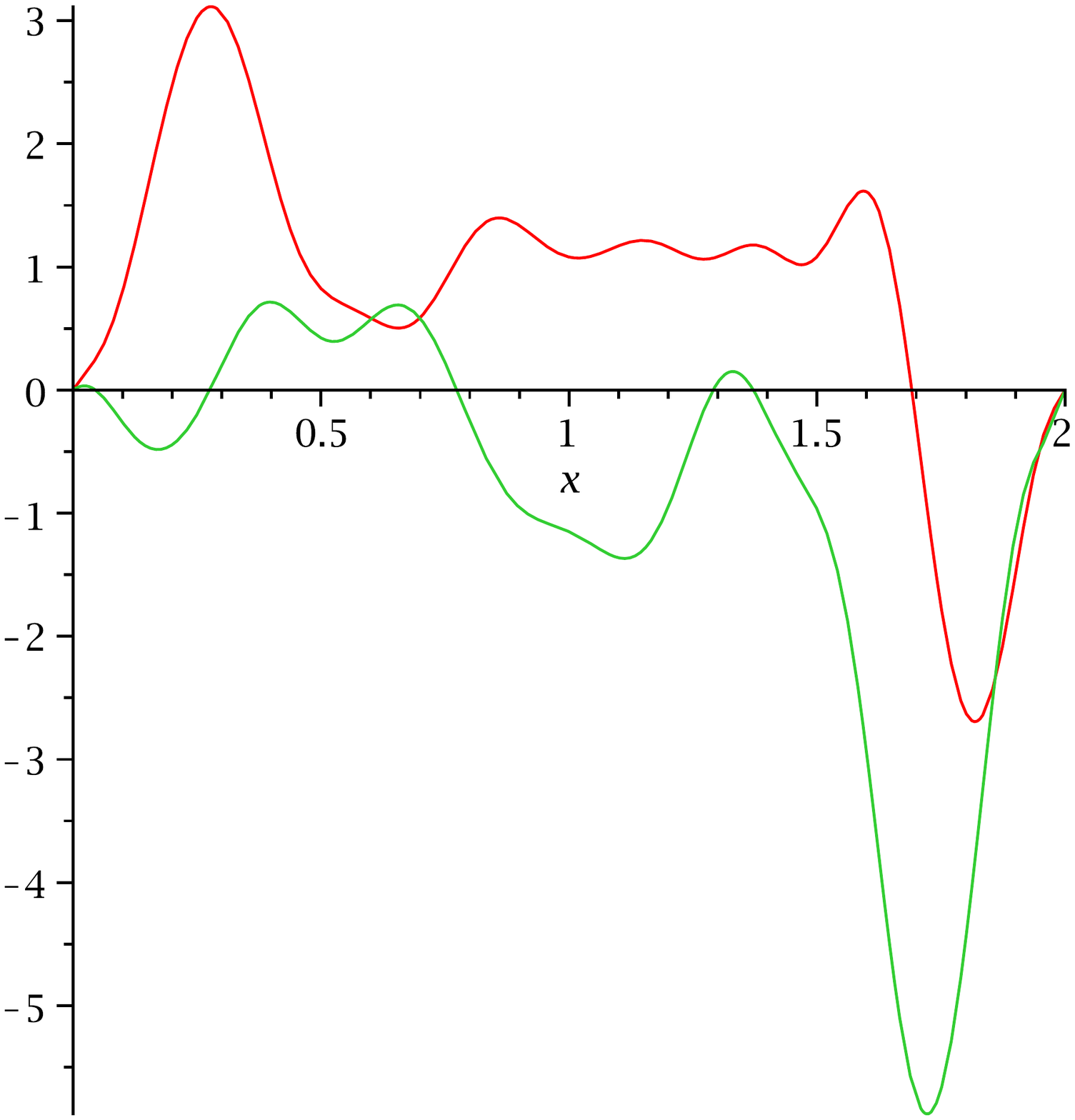}
\end{minipage}
\caption{A plot of the real and imaginary parts of the genus two theta functions
(\ref{fafactora})
for  $k=0,1,2$.}\label{h3all}
\end{figure}

\begin{figure}
\begin{minipage}[t]{0.3\textwidth}
\centering
\includegraphics[height=160pt]{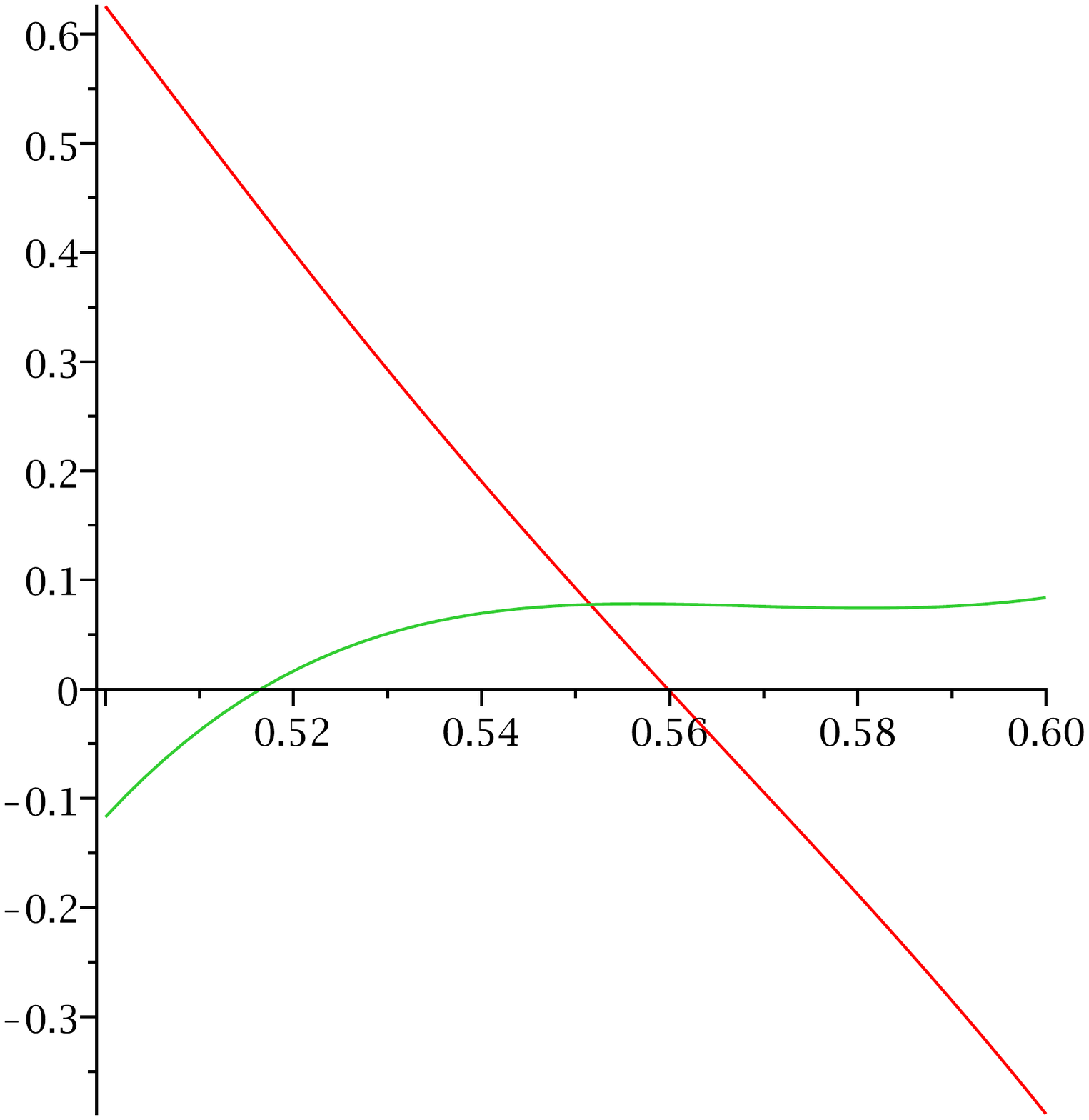}
\end{minipage}
\caption{An enlargement of the $k=0$ behaviour showing nonvanishing.}
\label{h3enlarged}
\end{figure}

There exists a rather nontrivial check of our results. We have
argued that the Hitchin constraint $\mathbf{H3}$ is automatically
satisfied by our construction. This means that each of the three
theta functions $\theta\left[\begin{matrix}0&0
\\ \frac{k}{3}&0 \end{matrix}\right]\left(\boldsymbol{ z};\tau
\right)$ of (\ref{fafactora}) (with $k=0,1,2$) should be
nonvanishing for $\lambda\in(0,2)$ and two of the three should
vanish at the endpoints. A sample check is shown in Figure
\ref{h3all} for $a=-12.3$, with Figure \ref{h3enlarged} showing an
enlarged portion of the $k=0$ curves to confirm the nonvanishing.
Such a numerical plot is the only means we know of for verifying
this condition.

\begin{figure}
\begin{center}
\includegraphics[height=200pt]{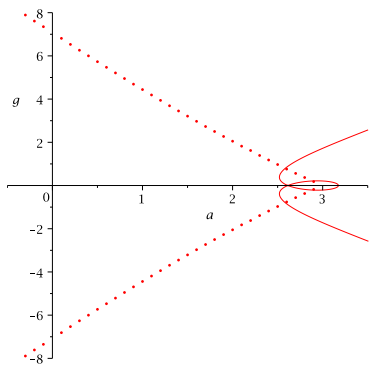}
\caption{The Genus $2$ curve covering elliptic curves}\label{asb3}
\end{center}
\end{figure}

Although our construction is generically in terms of a genus two
curve it may happen that this curve covers an elliptic curve. Such
occurs for the tetrahedral monopole. Shaska and V{\"o}lklein
\cite{sv04} describe when genus two curves give $2:1$ covers of
elliptic curves. For our curves these correspond to the axes $a=0$
($\texttt{D}_3$ symmetric monopoles) and $g=0$ and also to the
solid line given in Figure \ref{asb3}. We see that our curve
covers an elliptic curve at two further points for $0<a<3$ and
$g>0$ (and similarly for $g<0$). These points do not appear to be
otherwise special. For example, Sutcliffe studies the zeros of the
Higgs field and shows there is a point of positive $a$ for which
there is a `zero anti-zero' creation event \cite{sut96b}; both our
points are different from Sutcliffe's.

\section*{Acknowledgements}
We wish to thank T.P Northover for many fruitful discussions. His
programs \texttt{cyclepainter} and \texttt{extcurves} have been
used at many stages in this work.\footnote{These are available
from \texttt{http://gitorious.org/riemanncycles.}}

A.~D. is grateful for a Small Project Grant ($\#3606$) of the
University of Edinburgh which partially funded a visit to Kiev
where part of this work was developed. VZE is grateful to
Hanse-Wissenschaftskolleg (Institute for Advanced Study),
Delmenhorst, for a fellowship during which time the final version
of this paper was completed.

\providecommand{\bysame}{\leavevmode\hbox
to3em{\hrulefill}\thinspace}
\bibliographystyle{amsalpha}

\begin{thebibliography}{D{i}e91}

\bibitem[AF06]{af06}Abenda S., Fedorov, Yu. N. \emph{Closed
geodesics and billiards on quadrics related to elliptic KdV
solutions}, Lett. Math. Phys. {\bf 76} 111-134 (2006). {\tt
arXiv:nlin/0412034}.

\bibitem[Acc71]{acc71}Robert D. M. Accola, \emph{Vanishing Properties of Theta Functions
for Abelian Covers of Riemann Surfaces}, p7-18 in Advances in the
Theory of Riemann Surfaces: Proceedings of the 1969 Sony Brook
Conference, edited by L.V. Ahlfors, L. Bers, H.M. Farkas, R.C.
Gunning, I. Kra and H.E. Rauch (Princeton University Press 1971).


\bibitem[BE06]{bren06}
H.~W. Braden and V.~Z. Enolski, \emph{Remarks on the complex
geometry of
  3-monopole}, arXiv: math-ph/0601040, 2006.

\bibitem[BE09A]{bren09a}
\bysame, \emph{Finite-gap integration of the $SU(2)$ Bogomolny
equations},  Glasgow Math.J. {\bf 51 } (2009)  25--41.  arXiv:
\texttt{math-ph/ 0806.1807}

\bibitem[BE09B]{bren09}
\bysame, \emph{On the tetrahedrally symmetric monopole}, To appear
Commun. Math. Phys. arXiv: \texttt{math-ph/0908.3449}

\bibitem[BE10A]{bren10a}
\bysame, \emph{$SU(2)$-Monopoles, Curves
    with Symmetries and Ramanujan's Heritage}, {\em Matem.
    Sbornik} {\bf 201} (2010) 19--74.

\bibitem[BE10B]{bren10b}
\bysame, \emph{Some remarks on the
    Ercolani-Sinha construction of monopoles}, {\em Teor. Mat.
    Fiz}, 2010, {\em in press}.


\bibitem[BM88]{bm}{Jean-Beno{\^\i}t Bost and J F Mestre},
\emph{Moyenne {A}rithm\'etico-g\'eometrique et {P}\'eriodes des
{C}ourbes de genre 1 et 2},
     {Gaz.Math.S.M.F.} (1988), {36--64}.

\bibitem[Bra10]{bra10}H. W. Braden, \emph{Cyclic Monopoles, Affine Toda and
 Spectral Curves},
\texttt{arXiv:1002.1216}

\bibitem[CG81]{cg81} E. Corrigan and P. Goddard,
\emph{ An $n$ monopole solution with $4n-1$ degrees of freedom},
Commun. Math. Phys. \textbf{80} (1981), 575--587.

\bibitem[Cox84]{cox}{David Cox}, \emph{The {A}rithmetic-{G}eometric {M}ean of {G}auss},
     {L'Einsegnement Math\'ematique} \textbf{30} (1984), 275--330.

\bibitem[ES89]{es89}
N.~Ercolani and A.~Sinha, \emph{Monopoles and Baker Functions},
Commun. Math. Phys. \textbf{125} (1989), 385--416.

\bibitem[FK80]{fk80}
H. M. Farkas and I. Kra, {Riemann Surfaces}, Springer-Verlag, New
York, 1980.

\bibitem[Fay73]{fay73}
J.~D. Fay, {Theta functions on {R}iemann surfaces}, Lectures Notes
in
  Mathematics (Berlin), vol. 352, Springer, 1973.

\bibitem[Gau99]{gauss99} C. F. Gauss, \emph{Arithmetisch {G}eometrisches {M}ittel},
    In \emph{Werke}, Vol. 3, 361--432,
    Konigliche Gesellschaft der {W}issenschaft, {G}\"{o}ttingen,
    1799.
\bibitem[GH78]{gh}P. Griffiths and J. Harris,
 {{P}rinciples of {A}lgebraic {G}eometry}, Wiley, New York,
    1978.

\bibitem[Hit82]{hitchin82}
N.~J. Hitchin, \emph{Monopoles and {G}eodesics}, Commun. Math.
Phys. \textbf{83} (1982), 579--602.

\bibitem[Hit83]{hitchin83}
\bysame, \emph{On the {C}onstruction of {M}onopoles}, Commun.
Math. Phys. \textbf{89} (1983), 145--190.

\bibitem[Hit90]{hitchin90}
\bysame, \emph{Harmonic maps from a $2$-torus to the $3$-sphere},
 J. Differential Geom.  \textbf{31}  (1990), 627--710.

\bibitem[HMM95]{hmm95}
N. ~J. Hitchin, N. ~S. Manton and  M. ~K. Murray, \emph{Symmetric
monopoles}, Nonlinearity \textbf{8} (1995), 661--692.

\bibitem[HMR00]{hmr99}
C.~J. Houghton, N.~S. Manton, and N.~M. Rom{\~a}o, \emph{On the
constraints defining {BPS} monopoles}, Commun. Math. Phys.
\textbf{212} (2000), 219--243. arXiv: hep-th/9909168, 1999.

\bibitem[Hum01]{humbert} G. Humbert,
\emph{Sur la {T}ransformation {O}rdinaire des {F}onctions
{A}b\`eliennes},
    J. de Math. \textbf{7}(5) (1901).

\bibitem[KMMZ]{kmmz}V.A.Kazakov, A.Marshakov, J.A.Minahan, K.Zarembo,
\emph{Classical/quantum integrability in AdS/CFT} {\tt
arXiv:hep-th/0402207}


\bibitem[MS04]{ms04}Nicholas Manton and Paul Sutcliffe,
{Topological Solitons}, Cambridge University Press, Cambridge
2004.

\bibitem[Nah82]{nahm82} W.~Nahm, \emph{The construction of all
self-dual multimonopoles by the ADHM method}, in Monopoles in
Quantum Field Theory, edited by N.S.~Craigie, P.~Goddard and
W.~Nahm (World Scientific, Singapore 1982).


\bibitem[OR82]{or82} L.O'Raifeartaigh and S. Rouhani, \emph{Rings of monopoles with
discrete symmetry: explicit solution for n=3}, Phys. Lett.
\textbf{112B} (1982) 143.

\bibitem[Ric36]{richelot2}
    {F. Richelot}, \emph{Essai sur une {M}\'ethode {G}\'en\'erale pour {D}\'eterminer
    la {V}aleur des {I}nt\'egrales {U}ltra-elliptiques, {F}ond\'e sur les {T}ransformations
    {R}emarquables de ces {T}ranscendants}, {C. R. Acad. Sc.
    Paris} \textbf{2} (1836), 622--627.

\bibitem[Ric36]{richelot1}
    {F. Richelot}, \emph{De transforme {I}ntegralium {A}belianorum {P}rimi
    {O}rdinis {C}ommentation},
    {J. reine angew. Math.} \textbf{16} (1837), 221--341.

\bibitem[SV04]{sv04}Tanush Shaska and Helmut V{\"o}lklein,
\emph{Elliptic subfields and automorphisms of genus 2 function
              fields}, in
{Algebra, arithmetic and geometry with applications ({W}est
              {L}afayette, {IN}, 2000)}
703--723, {Springer} {Berlin}, 2004.

\bibitem[Sut97]{sut96b} Paul M. Sutcliffe, \emph{Cyclic
Monopoles}, Nucl.Phys. \textbf{B505} (1997) 517-539.
\texttt{arXiv:hep-th/9610030}


\end{thebibliography}

\end{document}